\documentclass{emulateapj}
\usepackage{apjfonts}
\usepackage{lscape}

\newcommand{\msigma}{$M_{\rm BH}-\sigma_{\star}$}
\newcommand{\mbh}{$M_{\rm BH}$}
\newcommand{\msun}{M$_{\odot}$}

\shorttitle{LAMP: OPTICAL H \& HE LINES}
\shortauthors{BENTZ ET AL.}

\begin{document}

\title{The Lick AGN Monitoring Project: Reverberation Mapping of
  Optical Hydrogen and Helium Recombination Lines}

\author{ 
Misty~C.~Bentz\altaffilmark{1,2},
Jonelle~L.~Walsh\altaffilmark{1},
Aaron~J.~Barth\altaffilmark{1},
Yuzuru~Yoshii\altaffilmark{3},
Jong-Hak~Woo\altaffilmark{4,5},
Xiaofeng~Wang\altaffilmark{6,7,8},
Tommaso~Treu\altaffilmark{9,10},
Carol~E.~Thornton\altaffilmark{1},
Rachel~A.~Street\altaffilmark{9,11},
Thea~N.~Steele\altaffilmark{6},
Jeffrey~M.~Silverman\altaffilmark{6},
Frank~J.~D.~Serduke\altaffilmark{6},
Yu~Sakata\altaffilmark{3,12},
Takeo~Minezaki\altaffilmark{3},
Matthew~A.~Malkan\altaffilmark{4},
Weidong~Li\altaffilmark{6},
Nicholas~Lee\altaffilmark{6,13},
Kyle~D.~Hiner\altaffilmark{14,15},
Marton~G.~Hidas\altaffilmark{9,11,16},
Jenny~E.~Greene\altaffilmark{17},
Elinor~L.~Gates\altaffilmark{18},
Mohan~Ganeshalingam\altaffilmark{6},
Alexei~V.~Filippenko\altaffilmark{6},
Gabriela~Canalizo\altaffilmark{14,15},
Vardha~Nicola~Bennert\altaffilmark{9,14}, and
Nairn~Baliber\altaffilmark{9,11}
}

\altaffiltext{1}{Department of Physics and Astronomy,
                 4129 Frederick Reines Hall,
                 University of California,
                 Irvine, CA 92697;
                 mbentz@uci.edu .} 

\altaffiltext{2}{Hubble Fellow.}

\altaffiltext{3}{Institute of Astronomy, 
                 School of Science, University of Tokyo, 
                 2-21-1 Osawa, Mitaka, Tokyo 181-0015, Japan.}

\altaffiltext{4}{Department of Physics and Astronomy, 
                 University of California, 
                 Los Angeles, CA 90024.}

\altaffiltext{5}{Astronomy Program, Department of Physics and
                  Astronomy, Seoul National University, Gwanak-gu,
                  Seoul 151-742, Korea}

\altaffiltext{6}{Department of Astronomy, 
                 University of California,
                 Berkeley, CA 94720-3411.}

\altaffiltext{7}{Physics Department and Tsinghua 
                  Center for Astrophysics (THCA), Tsinghua
                  University, Beijing, 100084, China.}

\altaffiltext{8}{Physics Department, 
                 Texas A\&M University, 
                 College Station, TX 77843-4242.}

\altaffiltext{9}{Physics Department, 
                 University of California, 
                 Santa Barbara, CA 93106.}

\altaffiltext{10}{Sloan Fellow, Packard Fellow.}

\altaffiltext{11}{Las Cumbres Observatory Global Telescope, 
                  6740 Cortona Dr. Ste. 102, 
                  Goleta, CA 93117.}


\altaffiltext{12}{Department of Astronomy,
                  School of Science, University of Tokyo, 
                  7-3-1 Hongo, Bunkyo-ku, Tokyo 113-0033, Japan.}

\altaffiltext{13}{Institute for Astronomy, 
                  2680 Woodlawn Dr., 
                  Honolulu, HI 96822.}

\altaffiltext{14}{Institute of Geophysics and Planetary Physics,
                 University of California,
                 Riverside, CA 92521.}

\altaffiltext{15}{Department of Physics and Astronomy,
                 University of California,
                 Riverside, CA 92521.}

\altaffiltext{16}{Sydney Institute for Astronomy,
                  School of Physics, 
                  The University of Sydney, 
                  NSW 2006, Australia.}

\altaffiltext{17}{Princeton University Observatory,
                 Princeton, NJ 08544;
                 Carnegie-Princeton Fellow.}

\altaffiltext{18}{Lick Observatory,
                 P.O. Box 85, 
                 Mount Hamilton, CA 95140.}

\begin{abstract}

We have recently completed a 64-night spectroscopic monitoring
campaign at the Lick Observatory 3-m Shane telescope with the aim of
measuring the masses of the black holes in 12 nearby ($z < 0.05$)
Seyfert~1 galaxies with expected masses in the range $\sim
10^6$--$10^7$~M$_{\odot}$ and also the well-studied nearby active
galactic nucleus (AGN) NGC\,5548.  Nine of the objects in the sample
(including NGC\,5548) showed optical variability of sufficient
strength during the monitoring campaign to allow for a time lag to be
measured between the continuum fluctuations and the response to these
fluctuations in the broad H$\beta$ emission, which we have previously
reported.  We present here the light curves for the H$\alpha$,
H$\gamma$, \ion{He}{2} $\lambda 4686$, and \ion{He}{1} $\lambda 5876$
emission lines and the time lags for the emission-line responses
relative to changes in the continuum flux.  Combining each
emission-line time lag with the measured width of the line in the
variable part of the spectrum, we determine a virial mass of the
central supermassive black hole from several independent emission
lines.  We find that the masses are generally consistent within the
uncertainties.  The time-lag response as a function of velocity across
the Balmer line profiles is examined for six of the AGNs.  We find
similar responses across all three Balmer lines for Arp\,151, which
shows a strongly asymmetric profile, and for SBS\,1116+583A and
NGC\,6814, which show a symmetric response about zero velocity.  For
the other three AGNs, the data quality is somewhat lower and the
velocity-resolved time-lag response is less clear.  Finally we compare
several trends seen in the dataset against the predictions from
photoionization calculations as presented by \citeauthor{korista04}.
We confirm several of their predictions, including an increase in
responsivity and a decrease in the mean time lag as the excitation and
ionization level for the species increases.  Specifically, we find the
time lags of the optical recombination lines to have weighted mean
ratios of $\tau{\rm (H\alpha)} : \tau{\rm (H\beta)} : \tau{\rm
  (H\gamma)} : \tau$(\ion{He}{1}) $ : \tau$(\ion{He}{2}) $ = 1.54 :
1.00 : 0.61 : 0.36 : 0.25$.  Further confirmation of photoionization
predictions for broad-line gas behavior will require additional
monitoring programs for these AGNs while they are in different
luminosity states.
\end{abstract}


\section{Introduction}

Active galactic nuclei (AGNs) are some of the most energetic objects
in the Universe, radiating at luminosities above
$10^{42}$\,erg\,s$^{-1}$, and yet their continuum emission is known to
vary on timescales as short as days.  The size constraints set by such
rapid variability mean that the extreme energy output of AGNs, often
comparable to or more than the energy output of all the stars in a
typical galaxy, must originate within a region whose size is $\sim
0.01$\,pc (approximately the size of our Solar System).  This large
amount of energy arising from such a small region is theorized to be
the result of gravitational accretion onto a supermassive black hole
(e.g., \citealt{rees84}).

For even the nearest AGNs, the region in which the continuum emission
arises is only microarcseconds in angular size and is therefore
unresolvable with current imaging detectors.  Dedicated monitoring
programs have instead taken advantage of the fast, and often dramatic,
variability of AGNs to completely revise our understanding of the
physical conditions present in the gas on these small scales.

Early monitoring programs with monthly sampling found that variations
in the broad emission lines promptly followed variations in the
continuum flux, putting an upper limit on the size of the broad-line
region (BLR) of only a light-month for typical Seyfert galaxies (e.g.,
NGC\,4151: \citealt{antonucci83}; Ark\,120: \citealt{peterson85}).
Especially in the case of Ark\,120, this upper limit was surprising,
as the size of the BLR was expected to be an order of magnitude
larger, based on photoionization models (e.g.,
\citealt{kwan81,ferland82}).  Higher temporal sampling has since
confirmed the size of the BLR for typical nearby Seyferts to be only a
few light-days.

In addition, densely sampled monitoring programs have discovered that
higher ionization lines respond more promptly (and more strongly) to
continuum variations than lower ionization lines (e.g.,
\citealt{clavel91}), indicating radial ionization stratification
throughout the BLR, contrary to the previous single-cloud models where
all emission lines were thought to arise from the same location.  More
recent models such as the ``locally optimally emitting cloud'' (LOC)
model \citep{baldwin95} predict ionization stratification as a natural
outcome.  In the LOC model, a range of cloud parameters is present in
the BLR and the emission that we happen to see as observers arises
from selection effects working within the BLR such that the majority
of the emission from a specific line will come from a location where
the parameters are most conducive to the production of that line.

A further discovery of monitoring programs is that the BLR appears to
be virialized; the distance to a specific region in the BLR is
inversely proportional to the square of the gas velocity in that
region.  This was first conclusively shown for the most well-studied
AGN, NGC\,5548 (\citealt{peterson99}; see also \citealt{krolik91}),
where $\tau \propto v^{-2}$, with $\tau$ the broad emission line time
lag relative to changes in the continuum flux (i.e., the BLR
light-crossing time), and $v$ the velocity width of the broad line.
Subsequent studies have also shown this to be true for several
additional AGNs (e.g., \citealt{onken02,kollatschny03}).  This
behavior is consistent with the fact that the BLR gas is under the
gravitational dominance of the central supermassive black hole, and so
the response of the BLR gas can be used to learn about the mass of the
black hole.

To date, black hole masses have been determined for some 44 AGNs
(\citealt{peterson04,peterson05,bentz09c}).  The most recent additions
come from the Lick AGN Monitoring Program (LAMP), a dedicated 64-night
spectroscopic monitoring campaign using the Lick Observatory 3-m Shane
telescope and supplemented by four small-aperture telescopes employed
in photometric monitoring.  First results from LAMP were presented by
\citet{bentz08} (hereafter Paper~I), followed by a full presentation
of the photometric light curves (\citealt{walsh09}; hereafter
Paper~II) and the H$\beta$ light curves and analysis
(\citealt{bentz09c}; hereafter Paper~III), and a re-examination of the
\msigma\ relationship for AGNs (\citealt{woo10}; hereafter Paper~IV).
In this work, we present the light curves and analysis for the
additional broad optical emission lines in the LAMP sample, namely
H$\alpha$, H$\gamma$, \ion{He}{2} $\lambda 4686$, and \ion{He}{1}
$\lambda 5876$.  We compare the results for these optical emission
lines with results from previous monitoring campaigns, as well as
with recent theoretical predictions of BLR behavior based on
photoionization models.

\section{Observations}

The LAMP sample of AGNs is comprised of 12 nearby ($z<0.05$) Seyfert~1
galaxies with single-epoch black hole mass estimates in the range
$\sim 10^6-10^7$~\msun, expected H$\beta$ lag times of $5-20$ days,
and relatively strong broad-line components to their H$\beta$ emission
lines.  In addition, we include NGC\,5548, the most well-studied AGN
with over a decade of densely sampled monitoring data and a
well-determined black hole mass from reverberation mapping of
$6.54^{+0.26}_{-0.25} \times 10^7$~M$_{\odot}$ (\citealt{bentz07}, and
references therein), for a total of 13 targets.

Each of the AGNs was monitored both photometrically (Johnson $B$ and
$V$ bands) and spectroscopically.  Details of the photometric
monitoring and data processing are presented in Paper~II.  In short,
four auxiliary telescopes were employed to monitor subsets of the LAMP
sample --- the 0.76-m robotic Katzman Automatic Imaging Telescope
(KAIT), the 2-m Multicolor Active Galactic Nuclei Monitoring
telescope, the Palomar 1.5-m telescope, and the 0.8-m Tenagra II
telescope.  The photometric monitoring began in early February 2008
and was increased to nightly monitoring on 2008 March 17 (UT, both
here and throughout), approximately two weeks before the onset of the
spectroscopic monitoring.  The images were reduced following standard
techniques and differential photometry was employed to determine the
brightness of the AGNs relative to stars within the field of view.
Absolute calibrations were set by observations of \citet{landolt92}
standard stars.  Finally, a simple galaxy disk model was determined
for each AGN host galaxy from images obtained on a night with good
seeing and clear skies.  The modeled disk flux determined to be within
the photometric aperture of the AGN was subtracted from the final AGN
light curves.  No correction has been attempted for the contribution
of bulge light, as the bulge and AGN point-spread function are
indistinguishable in the ground-based imaging.

Details of the spectroscopic monitoring and processing are presented
in Paper~III.  To summarize, spectroscopic monitoring was carried out
over 64 nights at the Lick Observatory 3-m Shane telescope between
2008 March 25 and June 1.  The red CCD of the Kast dual spectrograph
was employed with the 600~lines~mm$^{-1}$ grating (resulting in
spectral coverage over 4300--7100\,\AA), giving a nominal resolution
of 2.35\,\AA\,pix$^{-1}$ in the dispersion direction and
0\farcs78\,pix$^{-1}$ in the spatial direction.  A 4\arcsec-wide slit
was used and each target was observed at a fixed position angle.
IRAF\footnote{IRAF is distributed by the National Optical Astronomy
  Observatories, which are operated by the Association of Universities
  for Research in Astronomy, Inc., under cooperative agreement with
  the National Science Foundation.} was employed for the reduction of
the two-dimensional spectroscopic images and the extractions of the
one-dimensional spectra.  Flux calibrations were determined from
nightly spectra of standard stars, which typically included Feige\,34
and BD+28$^{\circ}$4211.  A final, internal calibration of the spectra
was accomplished using the spectral scaling algorithm of
\citet{vangroningen92}.  The algorithm scales the total flux of the
narrow [\ion{O}{3}] $\lambda\lambda$4959,~5007 doublet in each
spectrum to match the [\ion{O}{3}] flux in a reference spectrum
created from the mean of all the spectra for a given object.

As the [\ion{O}{3}] doublet is very close in wavelength to the
H$\beta$ emission line, the H$\beta$ line is the most accurately
calibrated broad line for each of our galaxies.  H$\alpha$, while
being much brighter than H$\beta$, is $\sim 1700$\,\AA\ redward of
H$\beta$ in the observed frames of these galaxies, and is near the red
edge of our spectroscopic coverage.  Unfortunately, there exists no
similar strong, unblended narrow emission line near H$\alpha$.  The
[\ion{S}{2}] doublet at $\lambda \lambda$6716,~6731 is fairly weak
and, at the typical redshifts for the LAMP AGNs ($z \approx 0.01$),
was often affected by the atmospheric B-band absorption at $\sim
6860-6890$\,\AA, making it an unacceptable choice for internal scaling
of the spectra.  Therefore, we have simply applied the scaling
determined for the H$\beta$ + [\ion{O}{3}] complex to the H$\alpha$
region as well, even though it may not be entirely accurate due to
aperture effects that can vary with wavelength (e.g., differential
atmospheric refraction, wavelength-dependent seeing).  The H$\gamma$
line is much less affected by these issues, as it is closer in
wavelength to the H$\beta$ line, although the data quality of
H$\gamma$ also occasionally suffers from being close to the blue edge
of our spectroscopic coverage.

\begin{figure*}
\plotone{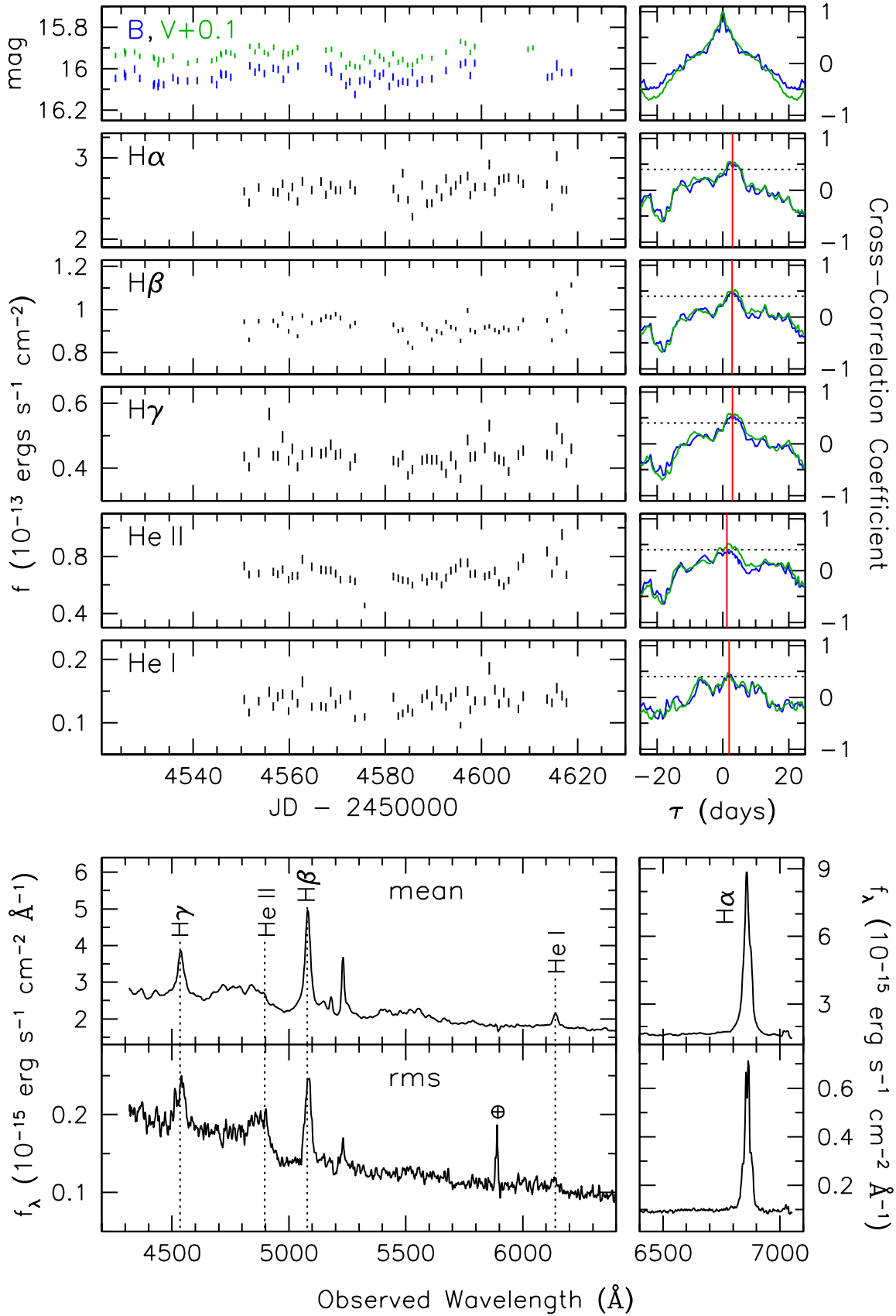}
\caption{{\it Top:} Photometric ($B$- and $V$-band) and spectroscopic
  light curves for the permitted broad optical emission lines in
  Mrk\,142.  The right panels show the cross-correlation functions
  versus the $B$ band (blue) and $V$ band (green).  For the
  photometric light curves, these are the auto-correlation
  functions. The horizontal dotted line in each CCF panel marks a
  significance of 0.4, and red vertical lines mark the locations of
  measured time lags, as listed in Table~\ref{table:tau}. {\it
    Bottom:} Mean and variable (rms) spectra of Mrk\,142.  The region
  around H$\alpha$ has been plotted separately on a different flux
  scale to allow for ease of viewing the weaker emission features at
  bluer wavelengths.  The spike at $5890$\,\AA\ is residual noise from
  the Na~I\,{\scriptsize\rm{D}\relax} night-sky line.}
\label{fig:mrk142}
\end{figure*}

\begin{figure*}
\plotone{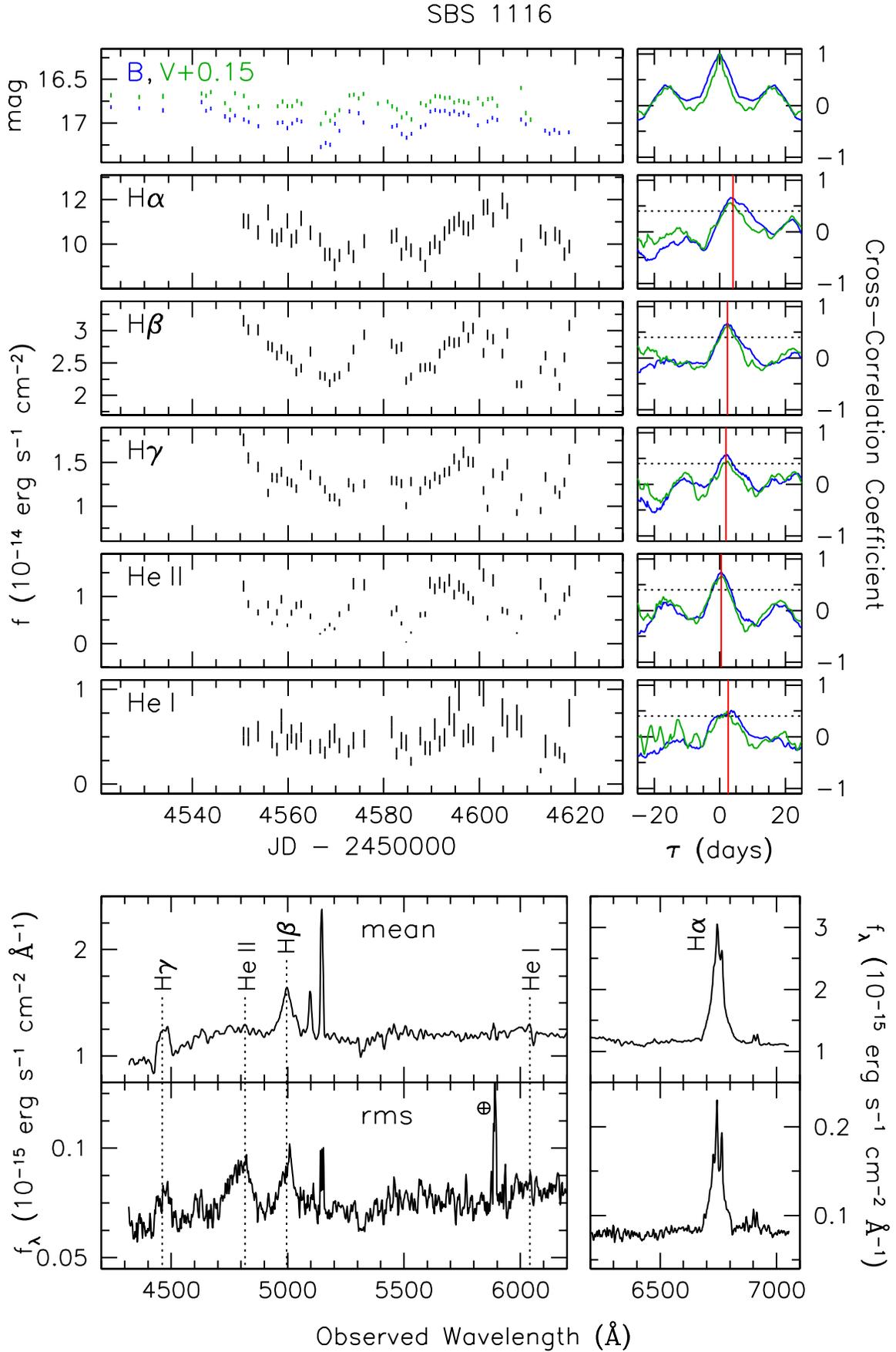}
\caption{Same as Figure~\ref{fig:mrk142}, but for SBS\,1116.}
\label{fig:sbs}
\end{figure*}

\begin{figure*}
\plotone{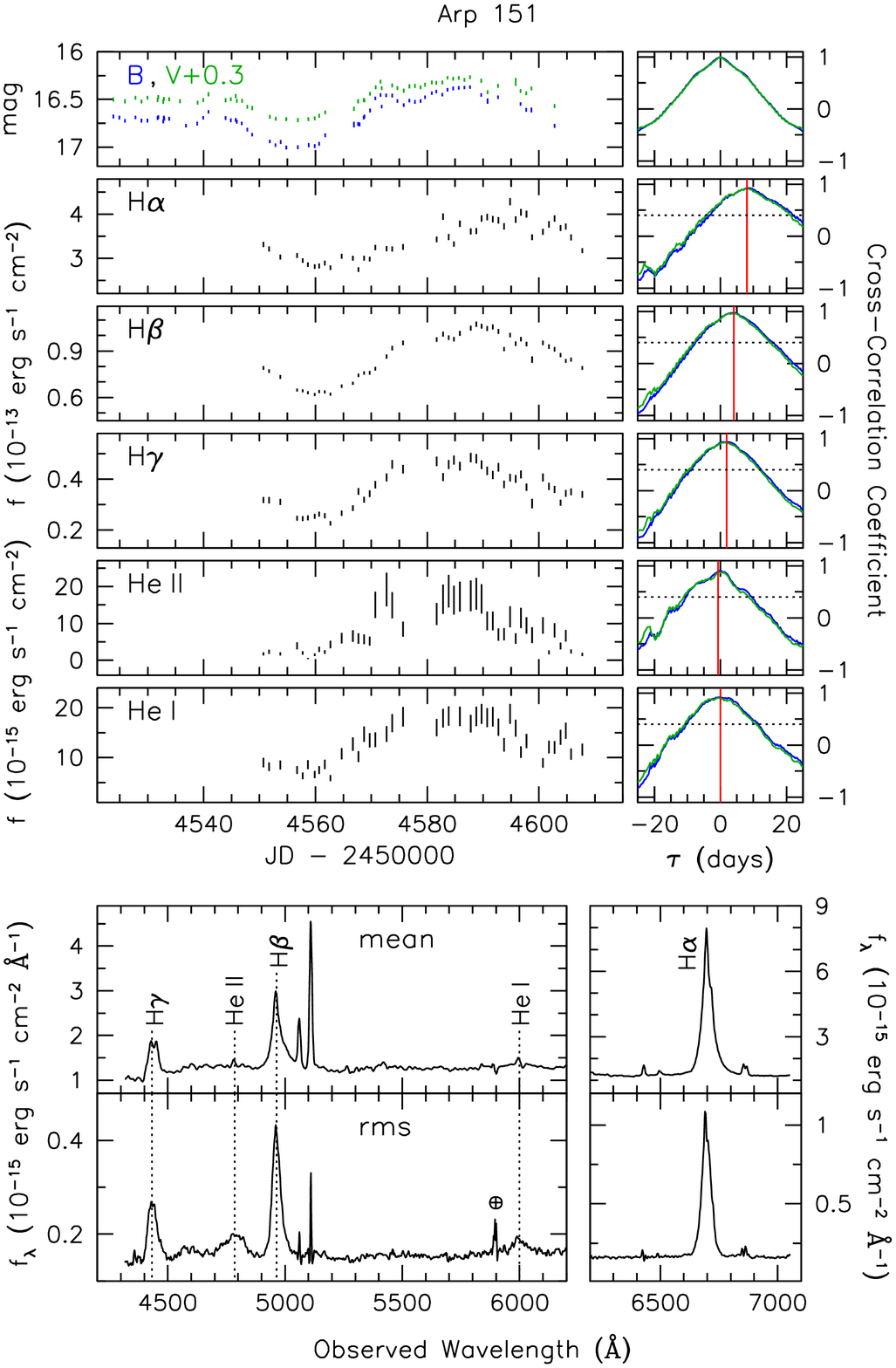}
\caption{Same as Figure~\ref{fig:mrk142}, but for Arp\,151.}
\label{fig:arp151}
\end{figure*}

\begin{figure*}
\plotone{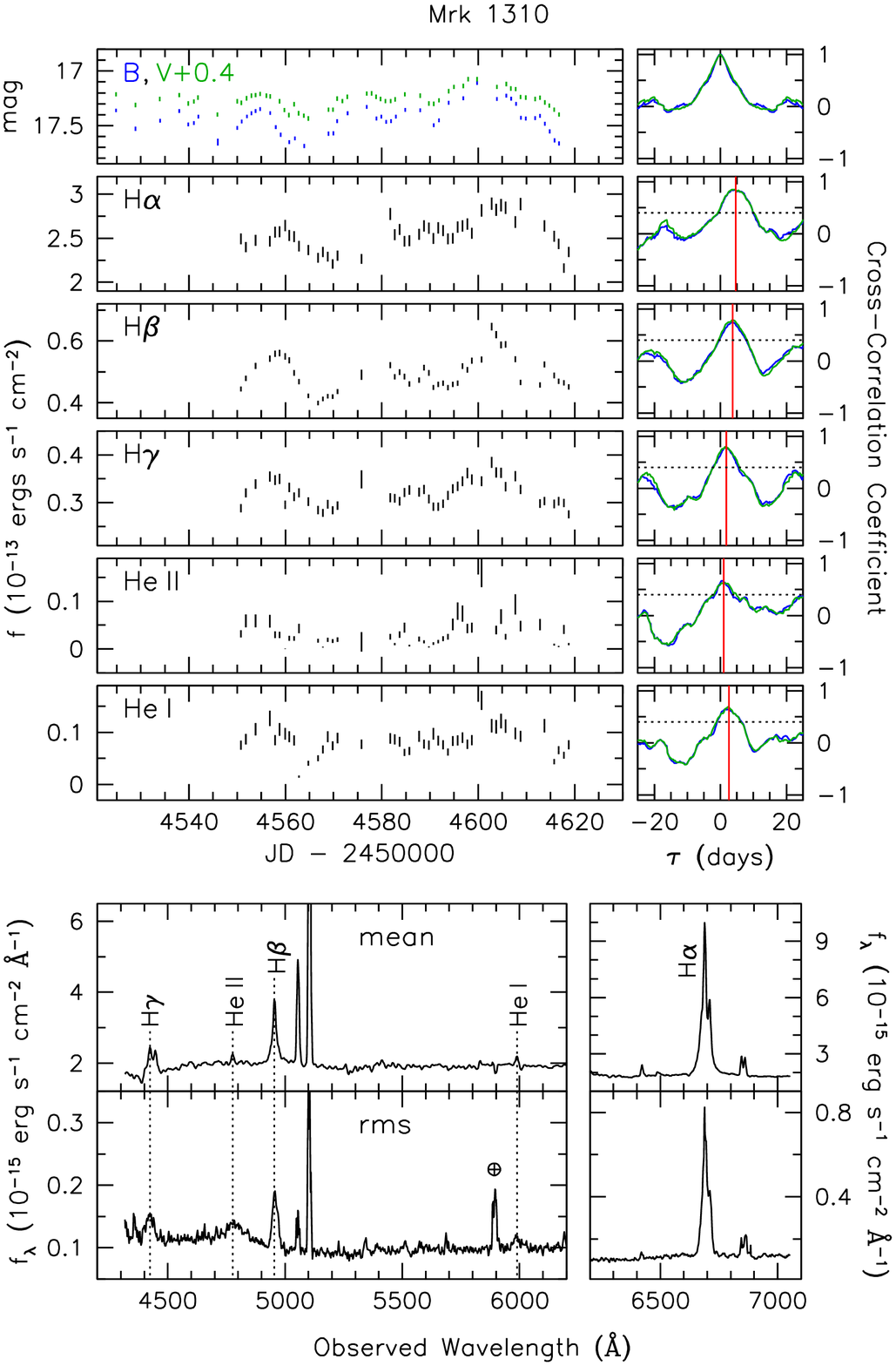}
\caption{Same as Figure~\ref{fig:mrk142}, but for Mrk\,1310.}
\label{fig:mrk1310}
\end{figure*}

\begin{figure*}
\plotone{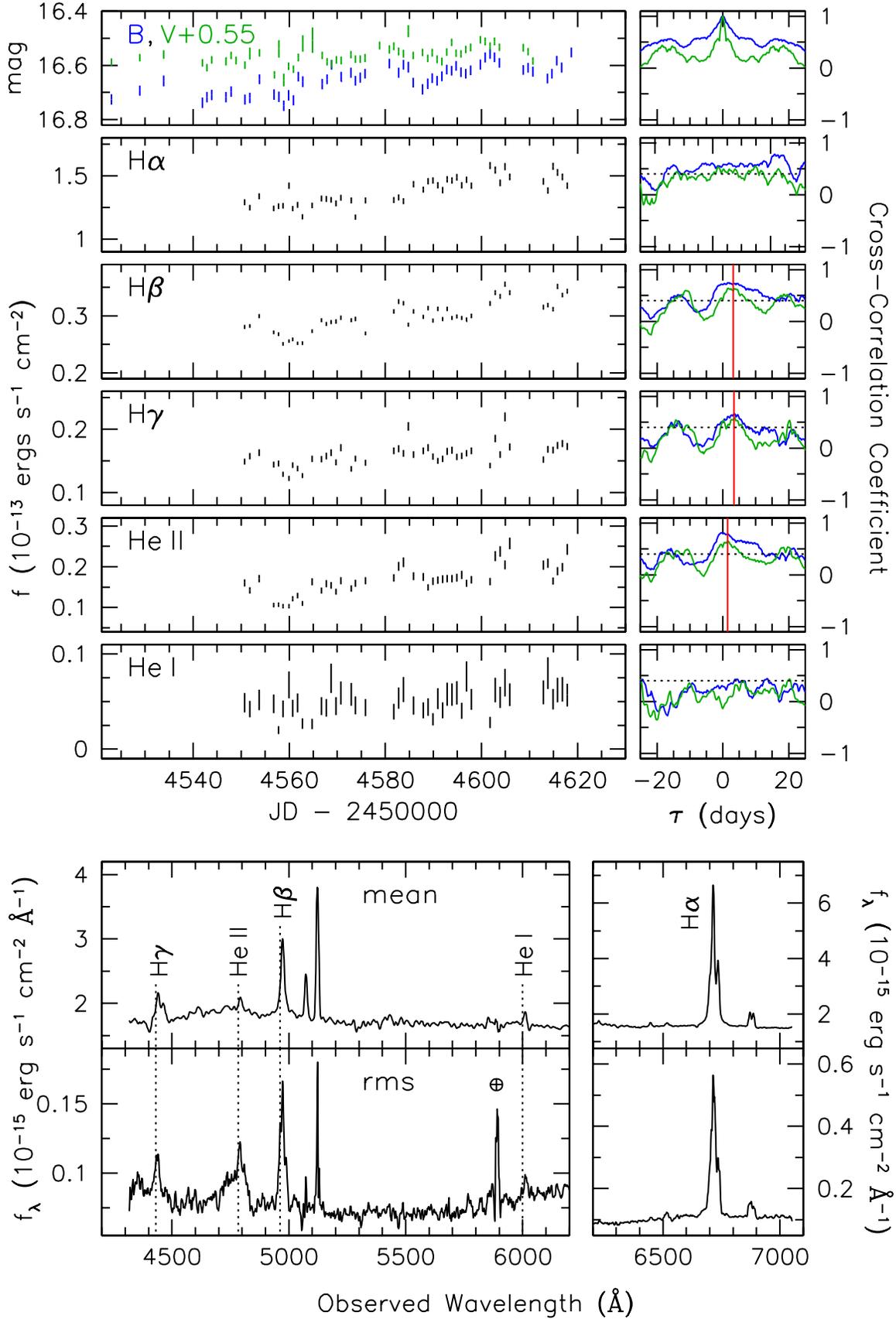}
\caption{Same as Figure~\ref{fig:mrk142}, but for Mrk\,202.}
\label{fig:mrk202}
\end{figure*}

\begin{figure*}
\plotone{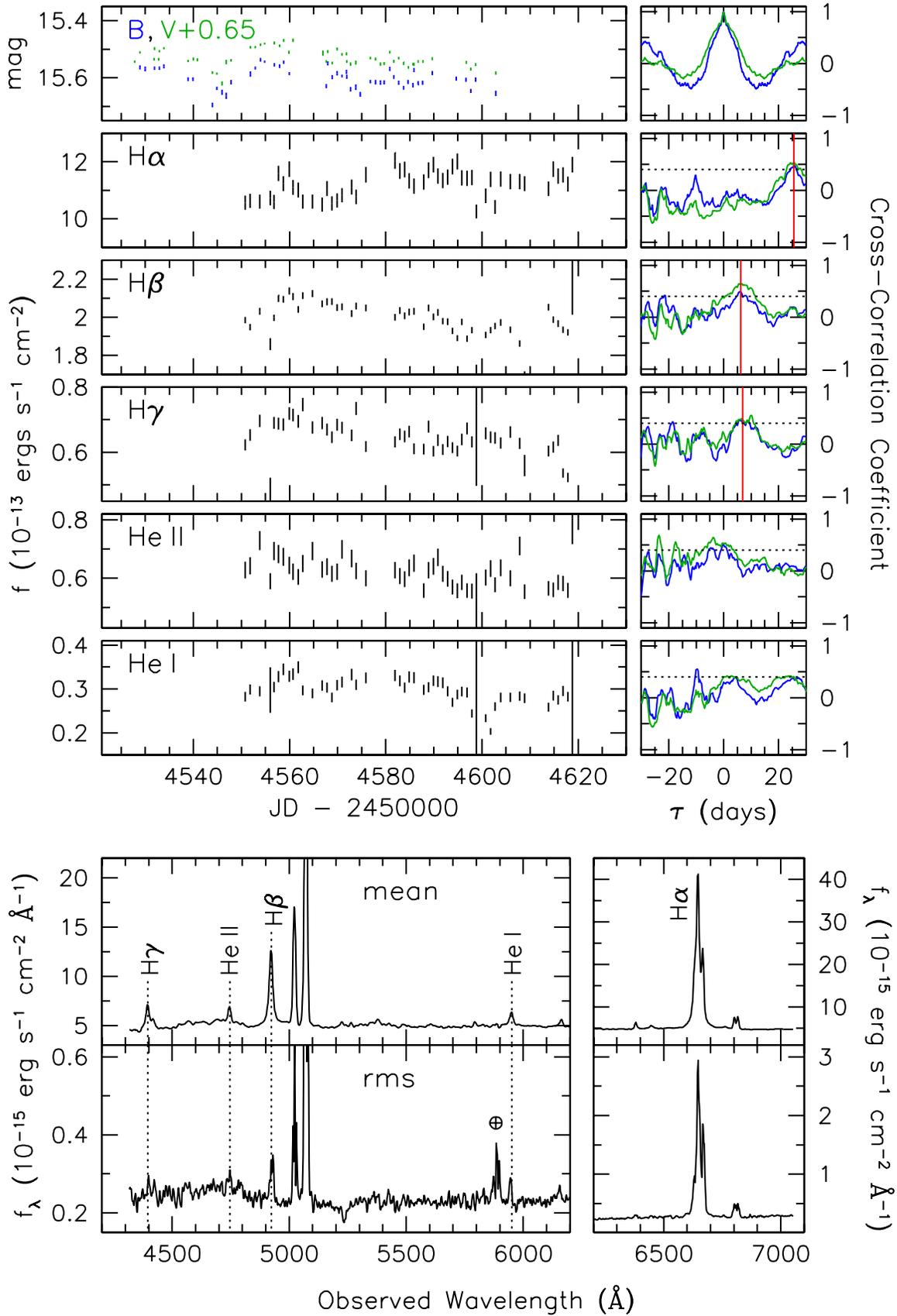}
\caption{Same as Figure~\ref{fig:mrk142}, but for NGC\,4253.}
\label{fig:mrk766}
\end{figure*}

\begin{figure*}
\plotone{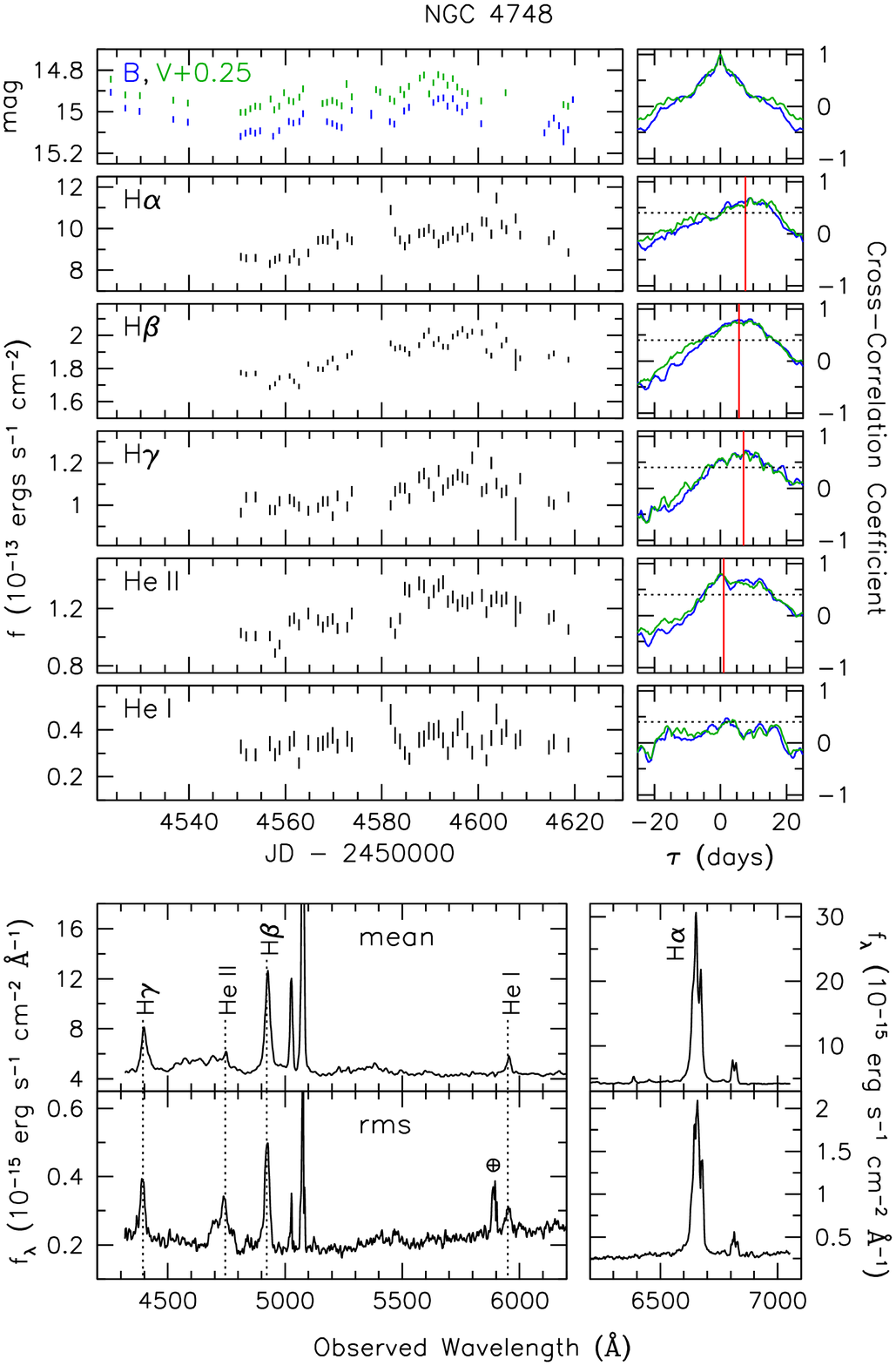}
\caption{Same as Figure~\ref{fig:mrk142}, but for NGC\,4748.}
\label{fig:n4748}
\end{figure*}

\begin{figure*}
\plotone{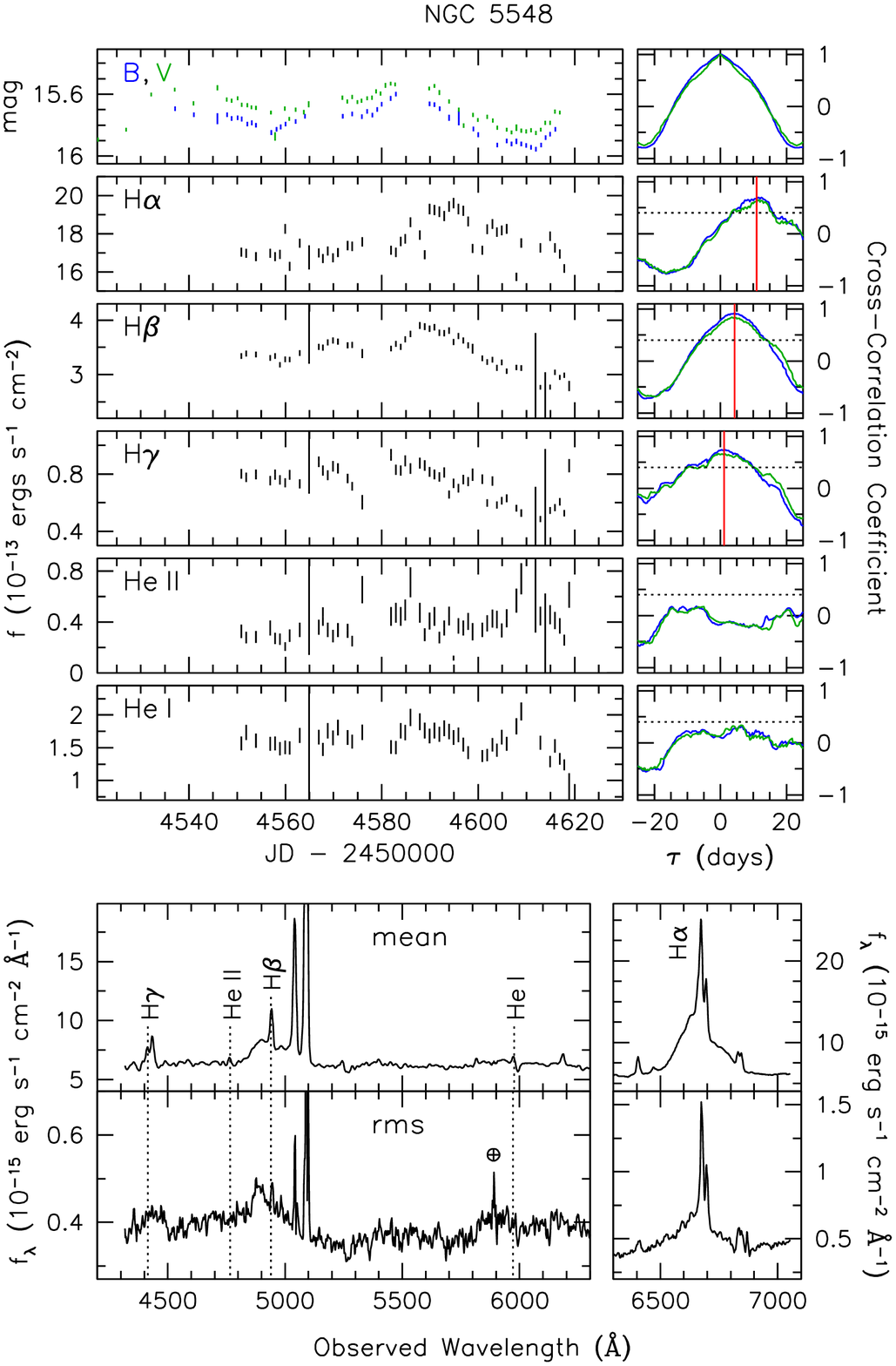}
\caption{Same as Figure~\ref{fig:mrk142}, but for NGC\,5548.}
\label{fig:n5548}
\end{figure*}

\begin{figure*}
\plotone{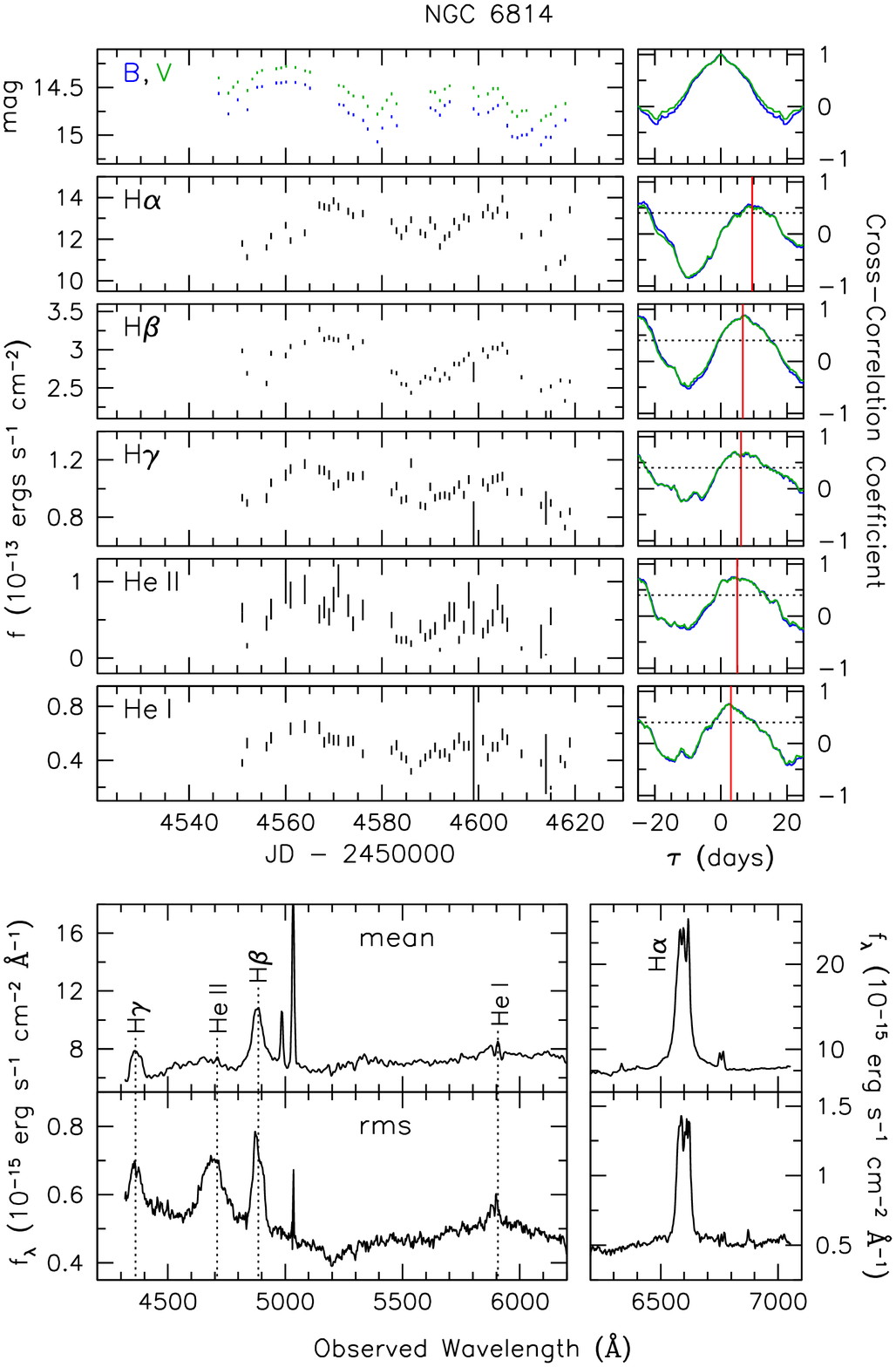}
\caption{Same as Figure~\ref{fig:mrk142}, but for NGC\,6814.}
\label{fig:n6814}
\end{figure*}

For each of the final, calibrated spectra, spectroscopic light curves
were measured by fitting a local, linear continuum under the broad
emission line and integrating the emission-line flux above the fitted
continuum.  This technique includes the flux contribution from the
narrow emission lines, but the contribution is just a constant flux
offset.  Table~\ref{table:objects} lists the nine LAMP targets for
which we were able to measure H$\beta$ time lags, and
Table~\ref{table:fluxwind} gives the continuum windows and line
integration limits for the broad, optical emission lines in each of
these nine AGNs.  Also listed in Table~\ref{table:fluxwind} are the
means and standard deviations of the emission-line fluxes.  In Paper
III, we discuss four objects in the LAMP sample which did not have
reliable H$\beta$ time lags (IC\,4218, MCG-06-30-15, Mrk\,290, and
IC\,1198).  We do not find any evidence for reliable time lags in any
of the additional optical broad-line light curves from any of these
four objects, and so we exclude them from further discussion.
Emission-line light curves for the nine LAMP AGNs are presented in
Tables~\ref{table:lc.mrk142}--\ref{table:lc.n6814} (we include here
the first five epochs of the light curves as a guide; the entirety of
the tables are available in the online journal).
Figures~\ref{fig:mrk142}--\ref{fig:n6814} display the $B$- and
$V$-band light curves, the emission-line light curves, and the mean
and root-mean-square (rms) spectra for each object.

Statistical properties of the light curves are listed in
Table~\ref{table:variability} along with the properties of the $B$-
and $V$-band light curves and the 5100\,\AA\ flux for comparison.
Column (1) lists the object, column (2) gives the feature, and column
(3) lists the number of measurements in each light curve.  Throughout
this analysis, we binned all photometric measurements within
0.1\,days. Columns (4) and (5) are the sampling intervals between data
points, measured as the mean and median, respectively. Column (6)
gives the mean fractional error, which is based on the comparison of
observations that are closely spaced in time.  The ``excess variance''
in column (7) is computed as

\begin{equation}
F_{\rm var} = \frac{\sqrt{\sigma^2 - \delta^2}}{\langle f \rangle},
\end{equation}

\noindent where $\sigma^2$ is the variance of the fluxes, $\delta^2$ is 
their mean-square uncertainty, and $\langle f \rangle$ is the mean of
the observed fluxes.  Finally, column (8) is the ratio of the maximum to
the minimum flux ($R_{\rm max}$) for each light curve.

\section{Analysis}
\subsection{Time-Series Analysis}

For each object, we determined the time lags for all of the broad
optical emission lines in the LAMP spectra (H$\alpha$, H$\beta$,
H$\gamma$, \ion{He}{2} $\lambda 4686$, and \ion{He}{1} $\lambda 5876$)
relative to the two continuum light curves ($B$ and $V$) measured from
the photometric monitoring.  In Paper~III, we describe in detail the
cross-correlation methods used for determining the time lags between
the continuum light curves and the broad emission-line light curves.
Here we give a brief summary for completeness.

Time lags were measured using the interpolation cross-correlation
function method of \citet{gaskell86} and \citet{gaskell87} with the
modifications described by \citet{white94}.  Cross-correlation
functions (CCFs) are characterized by the maximum cross-correlation
coefficient ($r_{\rm max}$), the time delay corresponding to the
location of $r_{\rm max}$ ($\tau_{\rm peak}$), and the centroid of the
points about the peak ($\tau_{\rm cent}$) above some threshold value,
typically $0.8r_{\rm max}$.  The uncertainties in the time lags were
determined using the Monte Carlo ``flux randomization/random subset
sampling'' method described by \citet{peterson98b,peterson04}, in
which the data points in each light curve are randomly sampled and
then randomly altered by a Gaussian deviation of the flux uncertainty.
The CCF is calculated for the sampled and modified light curves, and
$r_{\rm max}$, $\tau_{\rm cent}$, and $\tau_{\rm peak}$ are measured
and recorded.  A distribution of measurements is built up from 1000
realizations, and the means of the cross-correlation centroid
distribution and the cross-correlation peak distribution are taken to
be $\tau_{\rm cent}$ and $\tau_{\rm peak}$, respectively.  The
uncertainties on $\tau_{\rm cent}$ and $\tau_{\rm peak}$ are defined
such that 15.87\% of the realizations fall above and 15.87\% fall
below the range of uncertainties, which, for a Gaussian distribution,
would correspond to $\pm 1\sigma$.

Together with the photometric light curves and the broad emission-line
light curves in the top panels of
Figures~\ref{fig:mrk142}$-$\ref{fig:n6814}, we also show the
cross-correlation (auto-correlation) functions for the emission-line
(photometric) light curves.  In general, we find reliable time-lag
measurements for all three of the Balmer lines in the LAMP spectral
coverage.  In this case, we define ``reliable'' as those CCFs for
which (1) there is an obvious peak with $\tau \ge 0$, (2) the
correlations agree for both the $B$ and $V$ bands, and (3) $r_{\rm
  max} > 0.4$.  The CCFs for the He lines tend to be much noisier than
for the Balmer lines and more often fail our definition of
reliability.  For the reliable lag measurements, Table~\ref{table:tau}
lists measurements of $\tau_{\rm cent}$ and $\tau_{\rm peak}$ in the
observed and rest frame of each AGN for all the emission lines
compared to both the $B$- and $V$-band light curves (except for
Mrk\,202, where we list the ``unreliable'' measurements for H$\alpha$
for comparison with the rest of the sample).  In the following
discussion, we will give preference to lag measurements determined
relative to the $B$-band light curve, since the variations in the $B$
band are typically stronger than in the $V$ band (see Papers~II and
III for a discussion of this topic).

\subsection{Line-Width Measurement}

The mean and rms spectra for each of the nine AGNs examined here are
displayed in the bottom panels of
Figures~\ref{fig:mrk142}--\ref{fig:n6814}.  The rms spectra show the
standard deviation per spectroscopic pixel of all the individual
spectra relative to the mean spectrum for an object. Thus, the rms
spectra display the variable spectral components.

For each emission line with a measured and reliable time lag, the
width of the line was measured in the mean and the rms spectra.  The
helium lines appear as extremely low-level features in the mean
spectra, so only the rms line widths are tabulated when measurements
were possible.  The details of the techniques for measuring the line
widths and their uncertainties are described in Paper~III.  For each
line, we record the width as determined by the full-width at
half-maximum flux (FWHM) and the second moment of the line profile,
the line dispersion ($\sigma_{\rm line}$).  Each of the line-width
measurements has been corrected for the dispersion of the spectrograph
in the manner described in Paper~III.

In general, any constant spectral components, such as emission from
the host galaxy or narrow-line region (NLR), should disappear in the
rms spectra.  However, in practice, small errors in spectral
calibration and residual aperture effects caused by the combination of
a nonzero slit size, atmospheric seeing, and a spatially resolved NLR,
are often revealed by emission from constant components that appears
in the rms spectra.  Unfortunately, several of the objects in the LAMP
sample with strong narrow-line emission suffer from this problem in
the region of the H$\alpha$ line.  The typical correction for this
problem is to remove the narrow-line emission from each individual
spectrum in a consistent manner, where the total flux of each narrow
line is a constant throughout the spectroscopic campaign, before
creating the mean and rms spectra.  This method works well for the
narrow component of the H$\beta$ line and the narrow [\ion{O}{3}]
lines in the LAMP spectra, where the spectral calibration is most
accurate; however, it does not correct the problem of residual
narrow-line emission in the rms profile of H$\alpha$.  Attempts to
remove the narrow H$\alpha$ and [\ion{N}{2}] $\lambda\lambda 6548,
6583$ lines in a consistent manner from each individual spectrum
causes the residuals to be worsened in the rms spectra.  This
indicates that the spectral calibration is not completely accurate at
the wavelengths around H$\alpha$.

Rather than attempting to remove the narrow lines in a manner that is
not consistent from spectrum to spectrum (which would introduce
further biases into the line-width measurements for H$\alpha$), we
have revised the uncertainties in the H$\alpha$ rms line-width
measurements to compensate for any bias from residual narrow emission.
For each of the objects, the rms H$\alpha$ profile was interpolated
across in an attempt to exclude the narrow emission, and the line
width was measured and compared to the width from the uncorrected rms
spectrum.  For many of the objects, the interpolated line widths fell
within the uncertainties for the line width measurement.  For
SBS\,1116 and NGC\,5548, where this was not the case, the rms
H$\alpha$ line width uncertainty in the positive direction was
increased in quadrature by the difference of the corrected and
uncorrected line widths.  As residual narrow-line emission will always
tend to bias the line-width measurement toward smaller values, the
correction to the line-width uncertainties is asymmetric and only
affects the uncertainty in the positive (larger line width) direction.

A further complication appears upon examination of the H$\gamma$ line
in NGC\,5548.  This emission line is very close to the edge of the
spectroscopic coverage, and in its current low-luminosity state, the
broad lines in NGC\,5548 are extremely broad --- FWHM$ \approx
10,000$~km~s$^{-1}$ compared to only a few $\times 1000$~km~s$^{-1}$
for the other objects in the sample.  It is likely that the H$\gamma$
line is not fully covered by the spectral range in the LAMP dataset,
and as such both the mean time lag (which would only measure the
response of part of the emission line) and the line width of the
H$\gamma$ line in NGC\,5548 are suspect.

Table~\ref{table:width} lists the rest-frame broad-line widths and
their uncertainties.  The line-width measurements for H$\beta$ were
already presented in Paper~III, but we include them here for
comparison with the other emission lines.

\subsection{Black Hole Mass}

The mass of the putative black hole is determined from the equation

\begin{equation}
    M_{\rm BH} = f \frac{c \tau v^2}{G},
\label{eqn:mbh}
\end{equation}

\noindent where $\tau$ is the mean time delay for a specific emission
line, $v$ is the velocity width of the line, $c$ is the speed of
light, and $G$ is the gravitational constant.

The factor $f$ in the above equation is a scaling factor of order
unity that depends on the detailed geometry, kinematics, and emission
processes of the line-emitting region.  To date, the value of $f$ is
unknown, both for individual galaxies and the population average.
Instead of adopting a value for $f$ that is based on assuming a
specific model of the BLR, we adopt the scaling factor determined in
Paper~IV of $\langle f \rangle \approx 5.2 \pm 1.2$, which is the
value required to bring the AGN \msigma\ relationship into agreement
with the local, quiescent galaxy \msigma\ relationship.  This
particular value of the scaling factor is based on the union of the
LAMP sample and the sample previously considered by \citet{onken04},
and is consistent with the scaling factor determined by
\citet{onken04} ($\langle f \rangle = 5.5 \pm 1.8$), which has been
widely used in the literature and was used in Paper~III describing the
H$\beta$-based black hole mass derivations for this sample.  Although
our adopted value of $f$ is derived for the specific case of H$\beta$,
we will assume in the following analysis that the virial coefficient
is the same for the lines discussed here.  This choice is justified in
the absence of observational or theoretical arguments for a varying
$f$ within the broad, optical recombination lines.  As we will show,
the general agreement of virial products across the emission lines
considered here confirms that this choice is appropriate.

Following the findings of \citet{peterson04}, we use the combination
of $\tau_{\rm cent}$ and $\sigma_{\rm line}$(rms) to determine the
mass of the black hole in each object from each individual emission
line.  Table~\ref{table:mbh} lists the black hole mass calculated from
each individual broad emission line for the nine objects presented in
this work.  We list both the ``virial product,'' which assumes that
$f=1$, as well as the adopted black hole mass using the scaling factor
from Paper~IV.

\section{Results}

\subsection{Time Lag, Line Width, and $M_{\rm BH}$ Consistency}

\begin{figure}
\epsscale{1}
\plotone{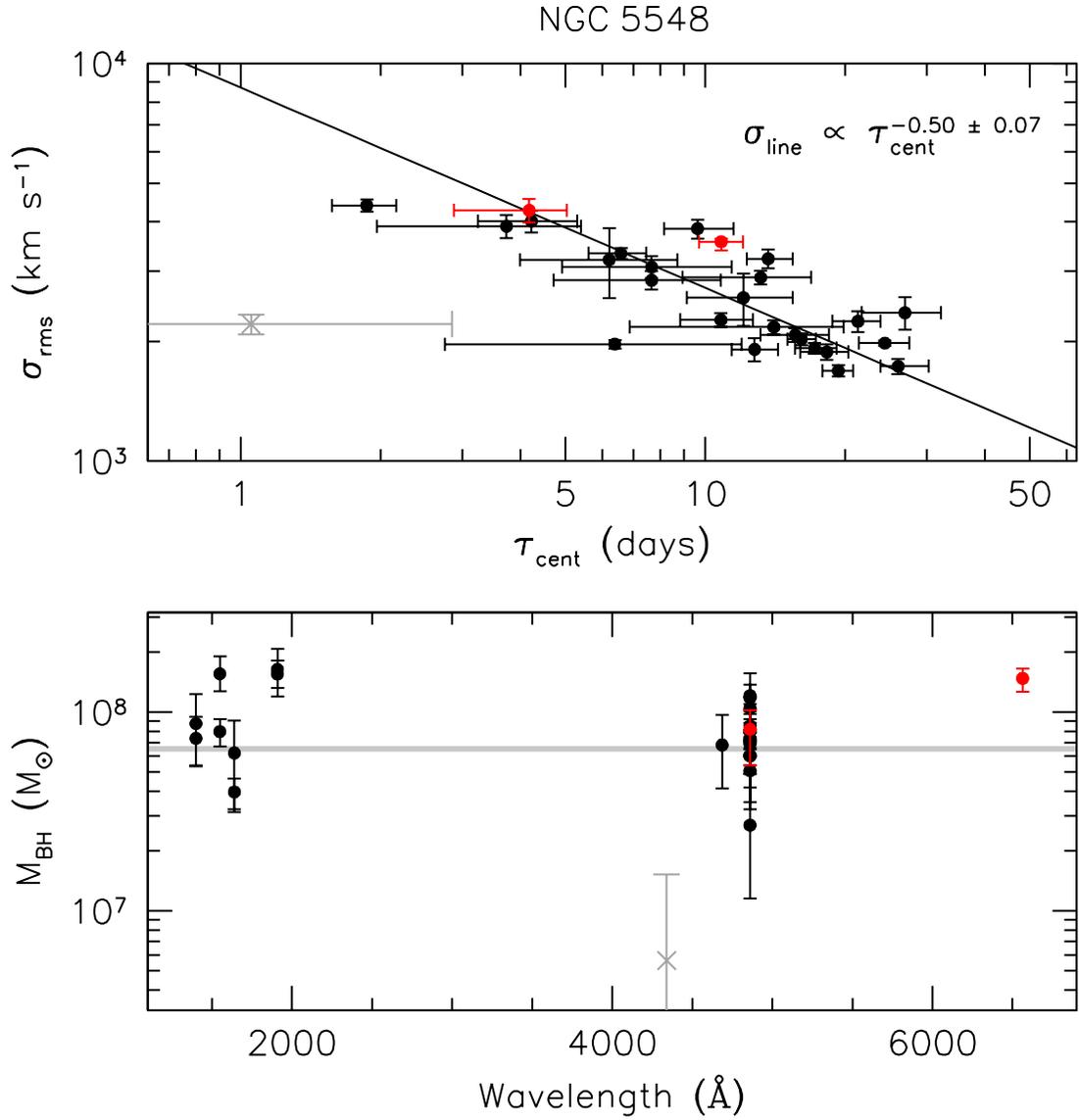}
\caption{{\it Top:} Broad emission-line width as a function of lag
  time for all reverberation results for NGC\,5548.  The solid line
  shows a least-squares fit to the data and has a power-law slope of
  $-0.50$, which is expected for a virial relationship.  Red points
  are the H$\alpha$ and H$\beta$ measurements presented in this work.
  The grey cross is for the H$\gamma$ measurement in this work, which
  is unreliable and was not included in the fit. {\it Bottom:} Black
  hole mass as a function of emission-line wavelength.  The gray band
  shows the $1\sigma$ range of the weighted average of the black
  hole mass based on H$\beta$ measurements.  Symbols are as described
  in the top panel.}
\label{fig:tl.n5548}
\end{figure}

NGC\,5548, with its many years of monitoring data, was the first AGN
known to show a virial relationship between time lag and line width,
indicative of the motion of BLR gas in a Keplerian potential
\citep{peterson99}.  In fact, this virial relationship holds for all
broad optical and ultraviolet emission lines for which a time lag
between the line and the continuum has been measured, as well as for
all the measurements of H$\beta$ that have been taken over some 15
years of monitoring campaigns (\citealt{bentz07}, and references
therein).  Several additional AGNs have since been shown to exhibit
virial behavior of their broad emission lines as well, such as
NGC\,3738 \citep{onken02} and Mrk\,110 \citep{kollatschny03}.
Figure~\ref{fig:tl.n5548} (top) shows the virial relationship between
lag and line width for all measured emission lines in NGC\,5548 as
determined by \citet{bentz07}, with the measurements of H$\alpha$ and
H$\beta$ from LAMP marked in red.  The H$\beta$ line measurements fall
directly on the fitted relationship, while the H$\alpha$ line
measurements lie above the relationship, but within the scatter.  Our
earlier concerns about the H$\gamma$ line-width measurements are
validated by the fact that the H$\gamma$ line measurements (marked
with a gray cross) lie far away from the locus of other emission-line
measurements and far below the fitted relationship.  A least-squares
fit \citep{mcarthur94} to all of the time-lag and line-width
measurements for NGC\,5548, including the uncertainties in both time
lag and line width, yields a slope of $-0.50 \pm 0.07$, in excellent
agreement with the expectation of $-0.5$ for a virial relationship.
Comparing all of the \mbh\ measurements from individual emission-line
reverberation results (Figure~\ref{fig:tl.n5548}, bottom), we see that
the mass based on H$\beta$ falls where expected, while that of
H$\alpha$ is somewhat high (albeit within the scatter).  The mass
measurement based on the H$\gamma$ measurements presented here is
extremely low and inconsistent with previous measurements.

\begin{figure*}
\plotone{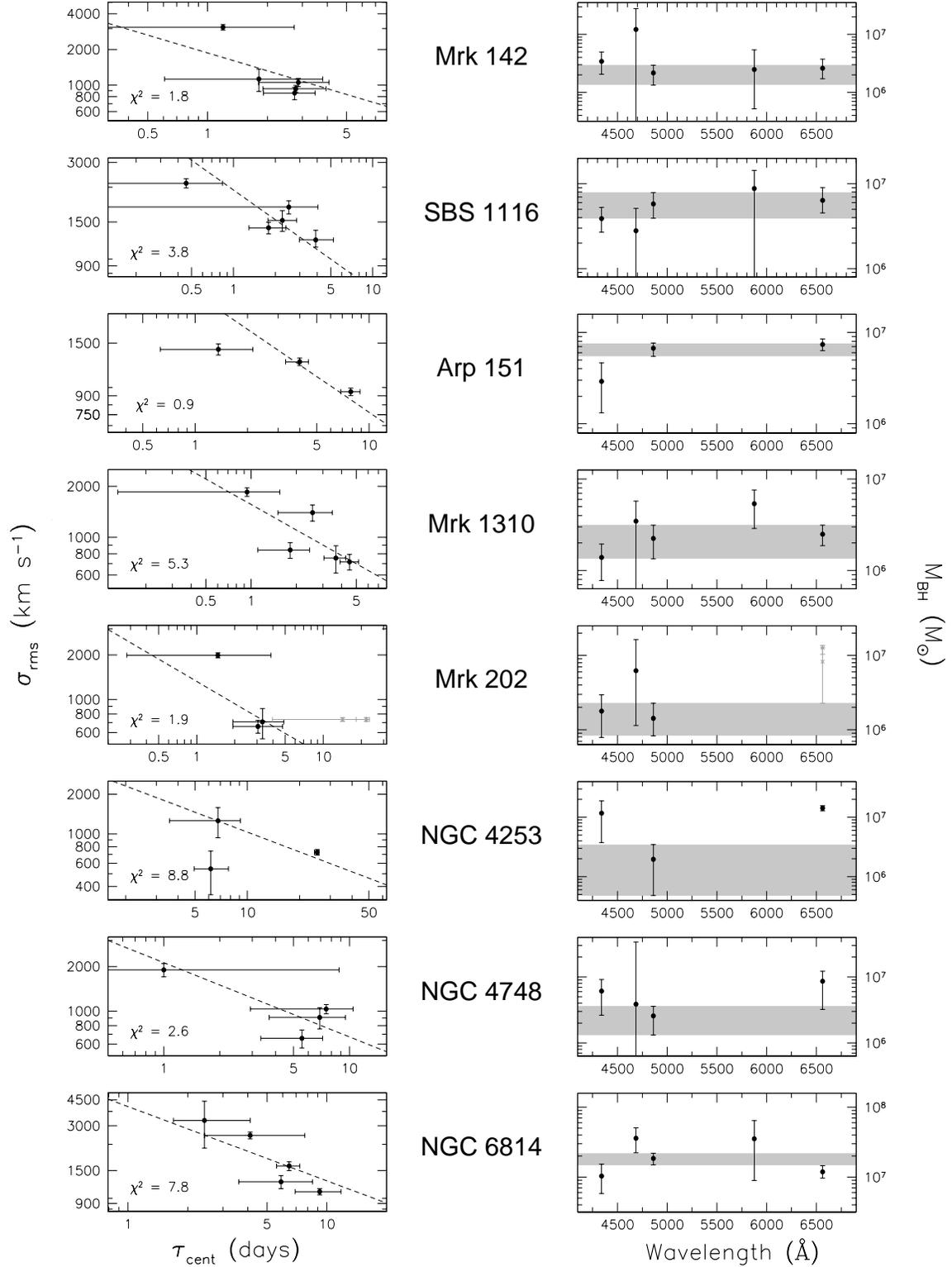}
\caption{{\it Left:} Broad emission-line width as a function of lag
  time for the other eight LAMP targets.  The dashed line shows a
  power-law slope of $-0.5$, which is expected for a virial
  relationship.  The grey crosses in the Mrk\,202 plot show the
  unreliable H$\alpha$ measurements for that object. {\it Right:}
  Black hole mass as a function of emission-line wavelength.  The gray
  band shows the $1\,\sigma$ range of black hole mass based on the
  measurements from the H$\beta$ line, which is the most accurately
  calibrated broad line in the LAMP spectra.  Symbols are as described
  for the left panels.}
\label{fig:tl}
\end{figure*}

Similar plots are presented for the other eight AGNs in
Figure~\ref{fig:tl}.  In the left panels we show the relationship
between time lag and line width for each AGN, and the right panels
show the black hole mass determined from each line.  With only a small
number of emission lines contributing to each plot, we do not attempt
to fit a power law to the relationship between lag time and line width
but instead display a power law of the form $v \propto \tau^{-0.5}$
with a dashed line to show the expected behavior for a virial
relationship.  In general, most objects are consistent with exhibiting
virial behavior. NGC\,4253 is perhaps the least consistent, but the
data quality for NGC\,4253 is rather low given the weak variability in
the AGN during our monitoring campaign.  In addition, we only have
measurements for three emission lines, so the inconsistency is not
surprising.  In the right panels, we show the black hole mass as a
function of emission-line wavelength.  Gray bands show the range
allowed by the $1 \sigma$ uncertainties in the black hole mass
determined from the H$\beta$ line (our most well-calibrated emission
line) for comparison with the black hole mass determinations from
other lines.  Again, the results are generally consistent within a
particular object except for NGC\,4253, but the black hole mass is
simply a combination of the line width and lag time, for which we have
somewhat poor measurements in this object.

\subsection{Velocity-Resolved Lag Measurements}

Reverberation mapping seeks to fully map out the response of the
line-emitting gas in the BLR as a function of both time and velocity.
In Paper~III we describe the expected behavior of three simple
kinematic models of the BLR (pure radial infall, ballistic outflow,
and circular orbits in a Keplerian potential) with the same geometric
and radiation parameters for each model (see Figure~10 of Paper~III
for a visual presentation of the expected responses across the line
profile).  In the case of circular orbits, the lag time as a function
of velocity is symmetric about the line center, and could even appear
flat across the emission line depending on the physical details of the
BLR.  Both infall and outflow show asymmetric behavior, with infall
having longer lags at blueshifted (negative) velocities and shorter
lags at redshifted (positive) velocities.  For outflow, the opposite
is expected.

In Paper~III, we present velocity-resolved lag times measured across
the H$\beta$ emission-line profile for six of our objects ---
SBS\,1116+583A, Arp\,151 (first presented in Paper~I), Mrk\,1310,
NGC\,4748, NGC\,5548, and NGC\,6814.  Using the same techniques
outlined there and in Section\,3.1 above, for these six objects we
divided both the H$\alpha$ and H$\gamma$ lines into four velocity bins
of equal variable flux and calculated the average lag time for each
velocity bin relative to the $B$-band light curve.  Attempts to do the
same for the \ion{He}{2} and \ion{He}{1} lines showed no difference in
lag measurements as a function of velocity.  We describe the results
for the Balmer lines for each of the individual objects below.

\paragraph{SBS\,1116+583A:}

\begin{figure*}
\epsscale{1}
\plotone{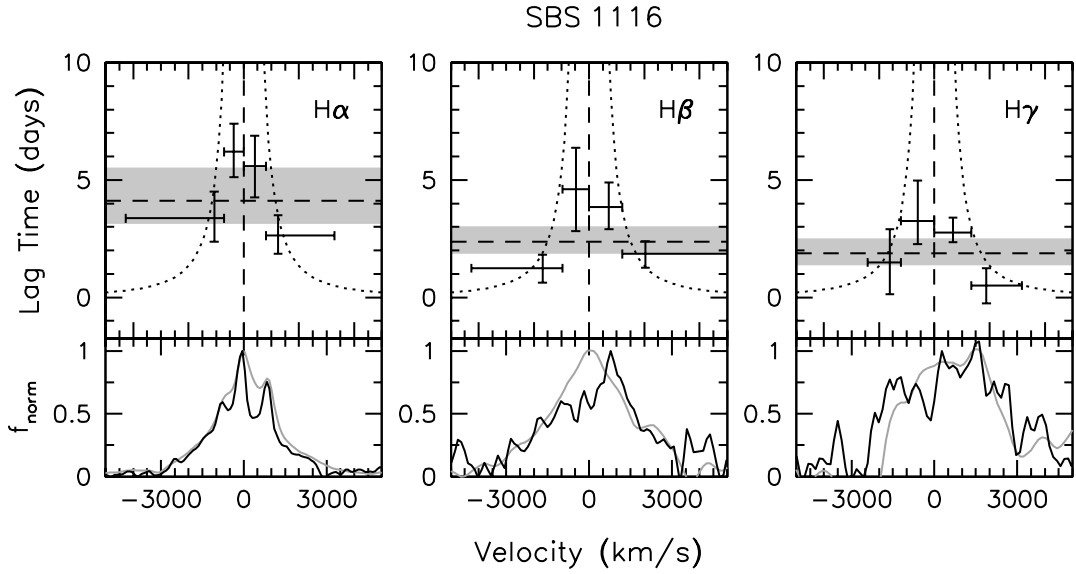}
\caption{{\it Top:} Velocity-resolved lag measurements for the broad
  Balmer-line emission in SBS\,1116.  The dashed line and grey band
  display the average time lag and the $1\sigma$ uncertainty,
  respectively, for each emission line.  The dotted curve in each
  panel is the Keplerian envelope for the adopted virial product based
  on the H$\beta$ time-lag and line-width measurements. {\it Bottom:}
  Black lines show the normalized variable (rms) Balmer emission-line
  profiles, while the grey lines show the normalized mean
  emission-line profiles for comparison. }
\label{fig:sbs.velres}
\end{figure*}

The H$\beta$ velocity-resolved lags for SBS\,1116 clearly show a
symmetric pattern about zero velocity, with shorter lags in the wings
and longer lags at the line center, indicative of circular orbits.
The same pattern can easily be seen in the H$\alpha$ and H$\gamma$
lines (see Figure~\ref{fig:sbs.velres}).  In addition, the H$\gamma$
line seems to show a double-peaked profile, an indication of flattened
geometry within the BLR gas.  Comparison of the H$\beta$ and H$\gamma$
lines reveals the possibility that the H$\beta$ variable emission is
also double peaked, with the stronger peak on the redshifted side as
is seen in H$\gamma$.  The strong residuals from the narrow H$\alpha$
and [\ion{N}{2}] lines superimposed on the variable broad H$\alpha$
emission do not allow a visual comparison of the variable H$\alpha$
profile with those of H$\beta$ and H$\gamma$.  Finally, the H$\gamma$
variable emission is somewhat blueshifted relative to the mean line
profile.

\paragraph{Arp\,151:}

\begin{figure*}
\plotone{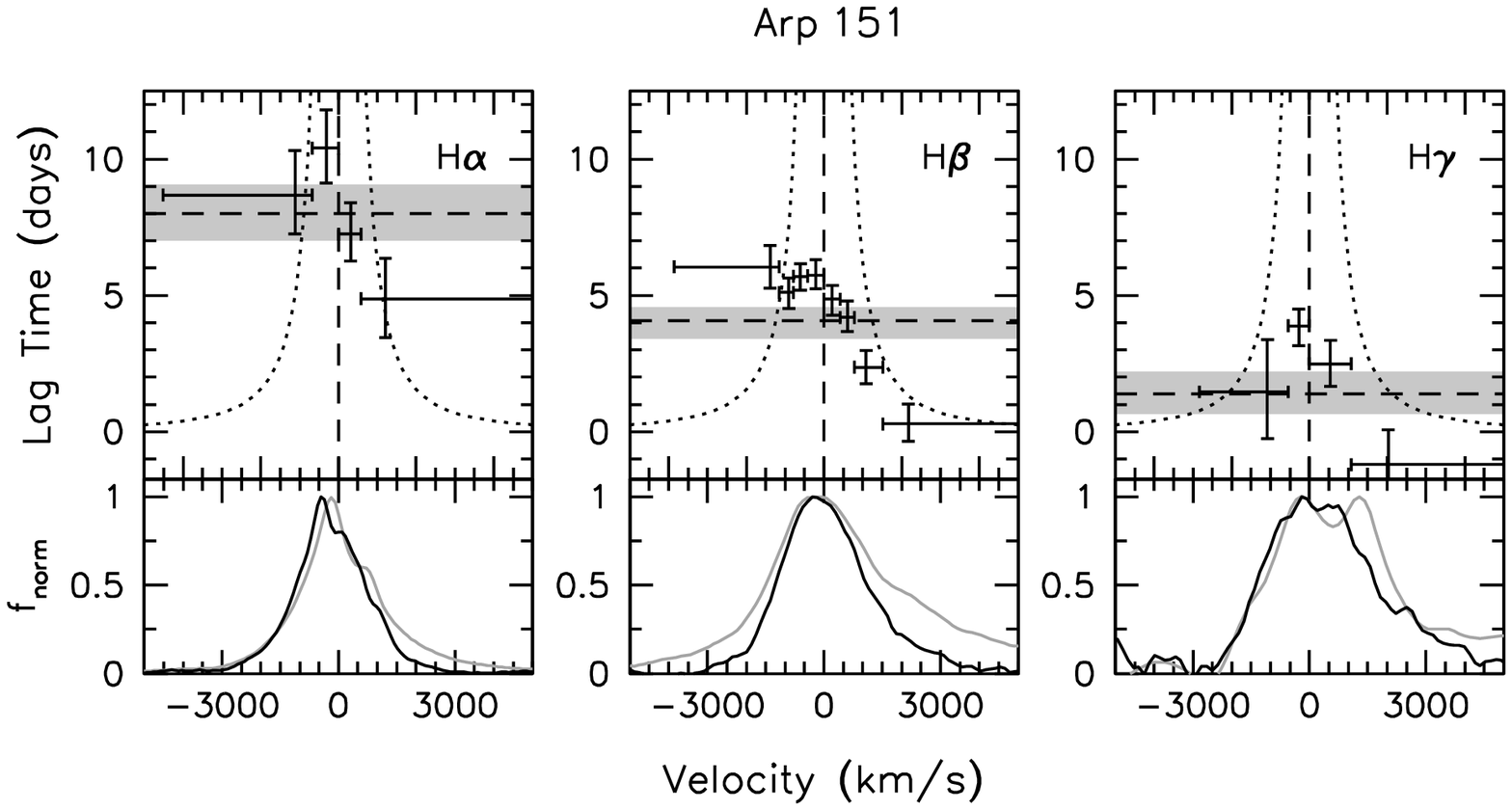}
\caption{Same as Figure~\ref{fig:sbs.velres}, but for Arp\,151.}
\label{fig:arp151.velres}
\end{figure*}

First described in Paper~I, the velocity-resolved H$\beta$ emission
has a strongly asymmetric lag behavior across the line profile.  The
blueshifted emission has long lag times that are higher than the total
mean lag time, while the redshifted emission drops off to almost zero
lag in the high-velocity gas in the wings.  This seems to imply simple
inflow in the BLR of Arp\,151.  Comparison with the
velocity-resolved lags measured for the H$\alpha$ and H$\gamma$ lines
in Figure~\ref{fig:arp151.velres} shows similar behavior, again
emphasizing the strong red-blue asymmetry in the emission-line
responses.  The variable line profiles appear to be single-peaked and
do not in general show a large velocity offset from the mean line
profile, although there seems to be excess emission in the red wing of
each line.

\paragraph{Mrk\,1310:}

\begin{figure*}
\plotone{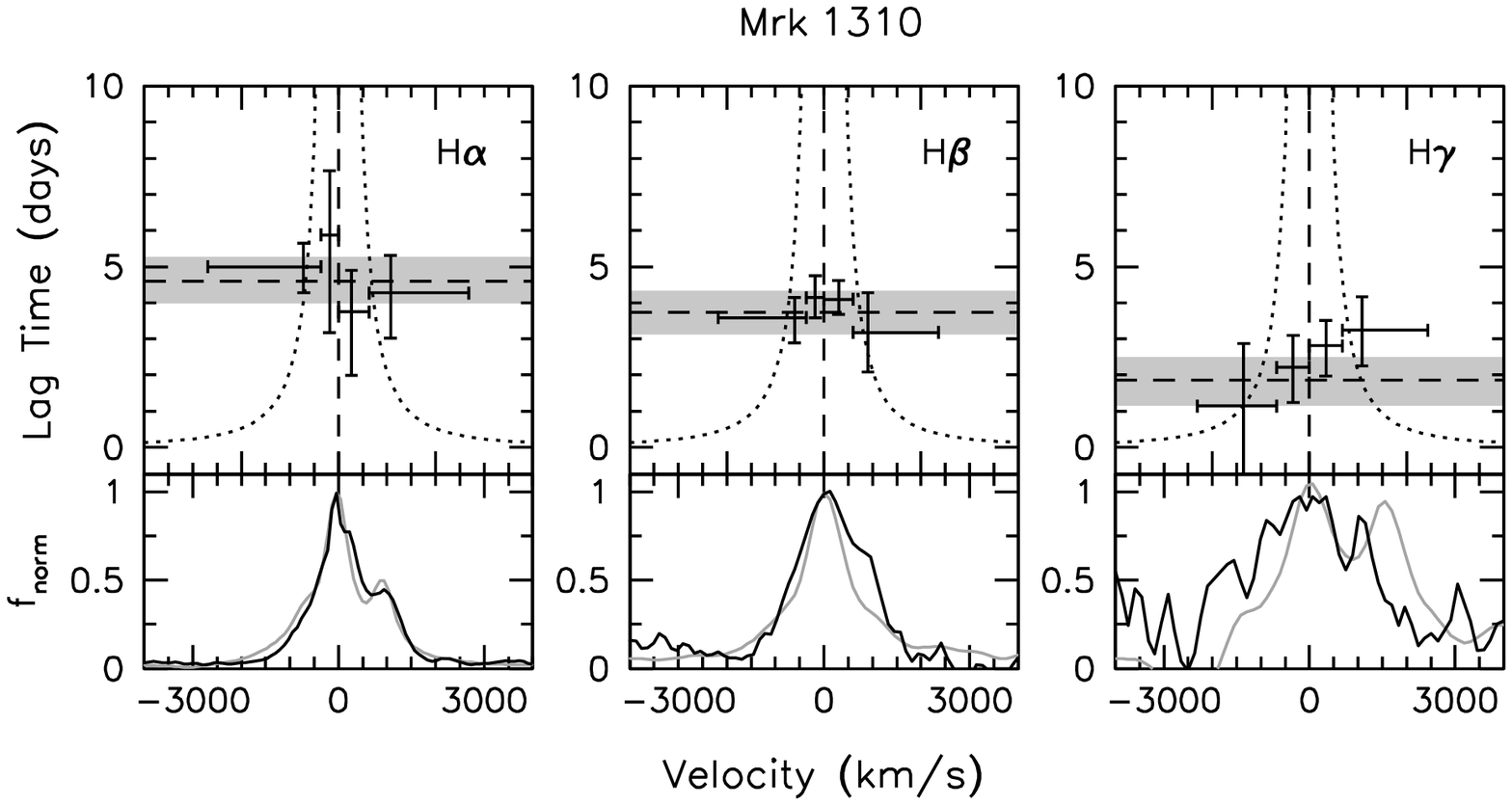}
\caption{Same as Figure~\ref{fig:sbs.velres}, but for Mrk\,1310.}
\label{fig:mrk1310.velres}
\end{figure*}

While the H$\beta$ velocity-resolved lags for Mrk\,1310 exhibit a
symmetric behavior about zero velocity, the H$\alpha$ and H$\gamma$
velocity-resolved structure is not so orderly, as shown in
Figure~\ref{fig:mrk1310.velres}.  Rather, the H$\gamma$ line seems to
exhibit evidence for outflow with a slight red-blue lag asymmetry, and
the variable line profile for H$\gamma$ is highly blueshifted from the
mean line profile.  The variable emission in H$\alpha$ and H$\beta$,
on the other hand, does not show a strong velocity shift relative to
the mean emission-line profiles.  More detailed analysis is clearly
needed to disentangle the BLR behavior of Mrk\,1310, but the
consistency with a flat response across the line profile may argue for
circular orbits in a Keplerian potential.

\paragraph{NGC\,4748:}

\begin{figure*}
\plotone{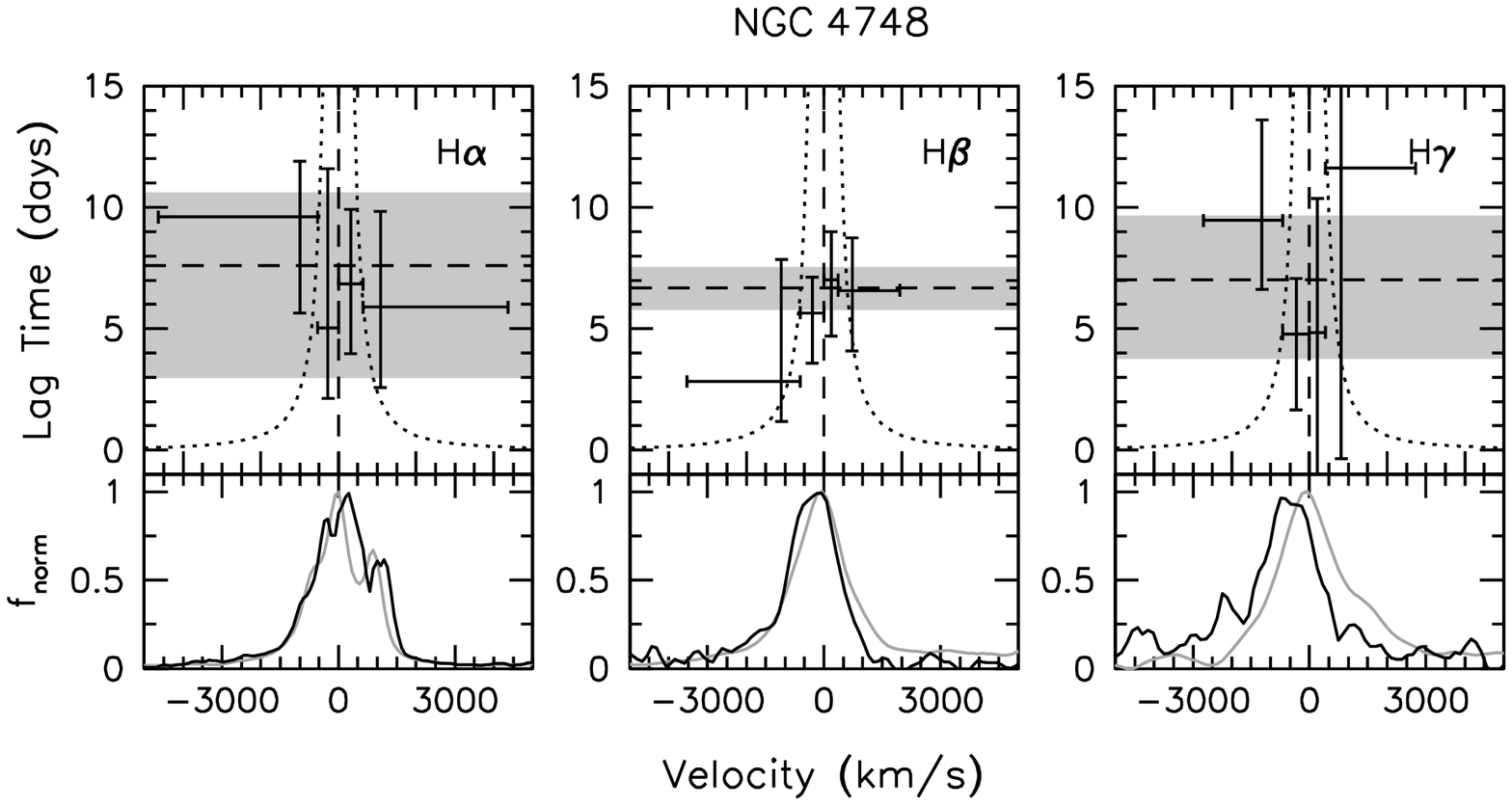}
\caption{Same as Figure~\ref{fig:sbs.velres}, but for NGC\,4748.}
\label{fig:n4748.velres}
\end{figure*}

As described in Paper~III, the H$\beta$ velocity-resolved lags in
NGC\,4748 (Figure~\ref{fig:n4748.velres}) may show evidence for
outflow with an asymmetric response about zero velocity.  The behavior
of the velocity-resolved lags within H$\alpha$ and H$\gamma$, however,
is not clear at all.  With the rather large uncertainties for this
object, the behavior is consistent with a flat response across the
emission lines, which would be consistent with circular orbits in a
Keplerian potential.  While the H$\alpha$ variable emission shows no
evidence for a large velocity shift, the H$\beta$ variable emission
shows a slight blueshift relative to the mean line profile, and the
H$\gamma$ variable emission is highly blueshifted.  There also appears
to be excess blue-wing emission in the variable flux of each Balmer
line.

\paragraph{NGC\,5548:}

\begin{figure*}
\plotone{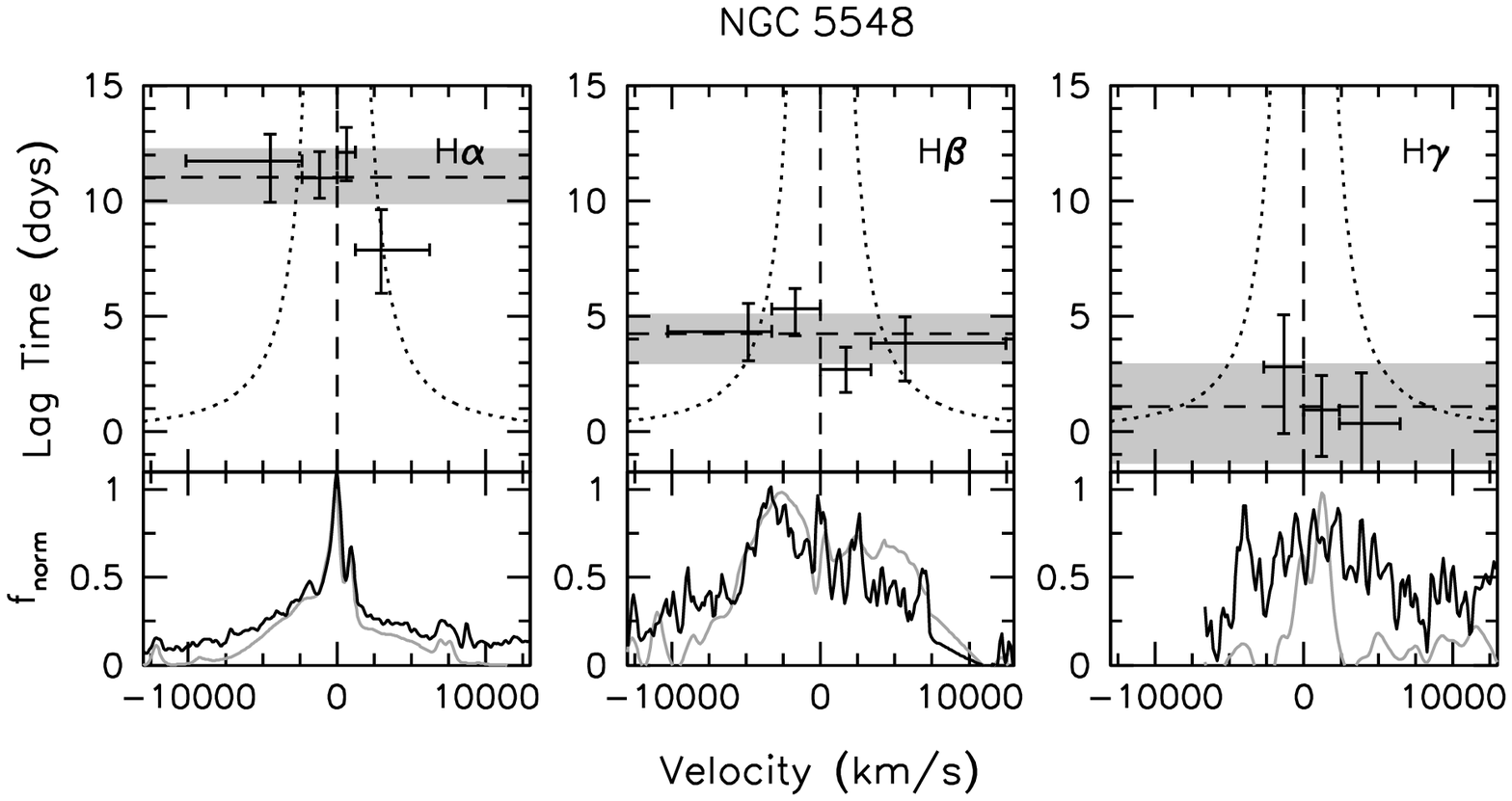}
\caption{Same as Figure~\ref{fig:sbs.velres}, but for NGC\,5548.}
\label{fig:n5548.velres}
\end{figure*}

The H$\beta$ velocity-resolved lag behavior for NGC\,5548 is not
particularly enlightening, and unfortunately, neither is that of
H$\alpha$ nor H$\gamma$ (see Figure~\ref{fig:n5548.velres}).  The
overall behavior appears to be consistent with a flat response across
the line profile, which could be consistent with circular orbits in a
Keplerian potential.  We have previously mentioned several reasons why
the H$\gamma$ line in this particular data set for NGC\,5548 may not
be reliable.  Given its shorter lag time, we would expect the
H$\gamma$ line to be broader than H$\alpha$ and H$\beta$.  Instead, it
appears that we may be missing a relatively large fraction of the
H$\gamma$ emission at the blue end, where the spectral coverage cuts
off.  For this reason as well as those previously mentioned, we will
classify the H$\gamma$ measurements included in this work as
``unreliable.''

\paragraph{NGC\,6814:}

\begin{figure*}
\plotone{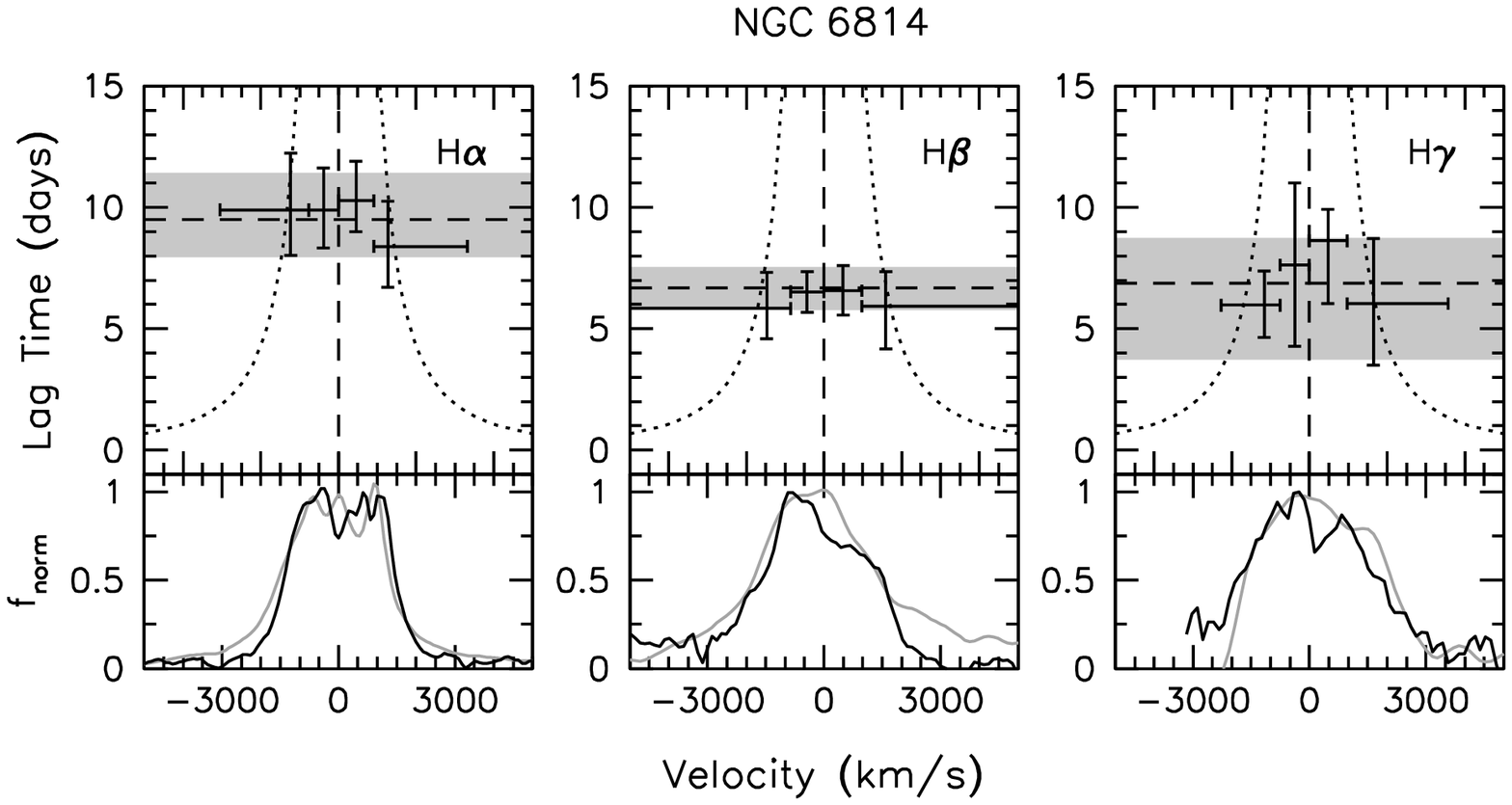}
\caption{Same as Figure~\ref{fig:sbs.velres}, but for NGC\,6814.}
\label{fig:n6814.velres}
\end{figure*}

Similar to SBS\,1116, the velocity-resolved lags across the H$\beta$
emission profile of NGC\,6814 show a symmetric behavior.  This
symmetric behavior is also seen in the H$\alpha$ and H$\gamma$ lines
(see Figure~\ref{fig:n6814.velres}).  All three emission lines seem to
have a double-peaked profile shape in the variable emission, possibly
indicative of a disk-like geometry in the BLR.  There does not appear
to be any significant velocity shift in the variable emission compared
to the mean line profiles, and the widths of the variable and mean
profiles are very similar, demonstrating that the full range of gas
giving rise to the integrated line flux is responding to changes in
the continuum flux.

\subsection{Comparison with Photoionization Predictions}

With the large number of optical emission lines ($\sim 5$) for which
we have carried out a reverberation-mapping analysis in each object,
we are able to examine and compare the behavior of trends that are
exhibited among emission lines with predictions from photoionization
calculations of BLR-like gas.  Here, we focus on the specific
predictions for the optical recombination lines presented by
\citet{korista04}.  Their predictions are based on a grid of
photoionization calculations, originally presented by
\citet{korista00}, and generated with CLOUDY \citep{ferland98} to
model the broad UV emission lines in NGC\,5548.

\begin{figure*}
\plottwo{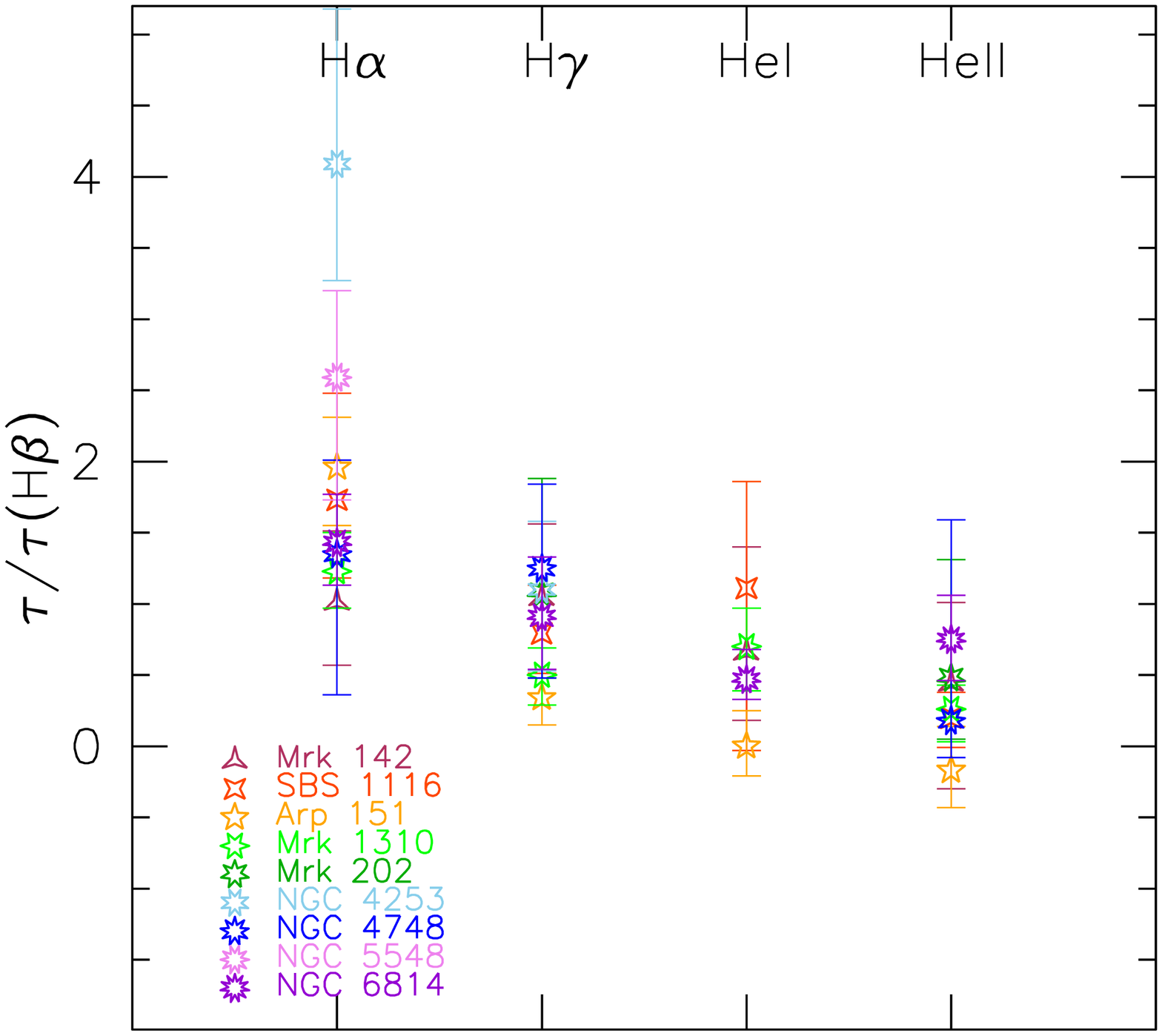}{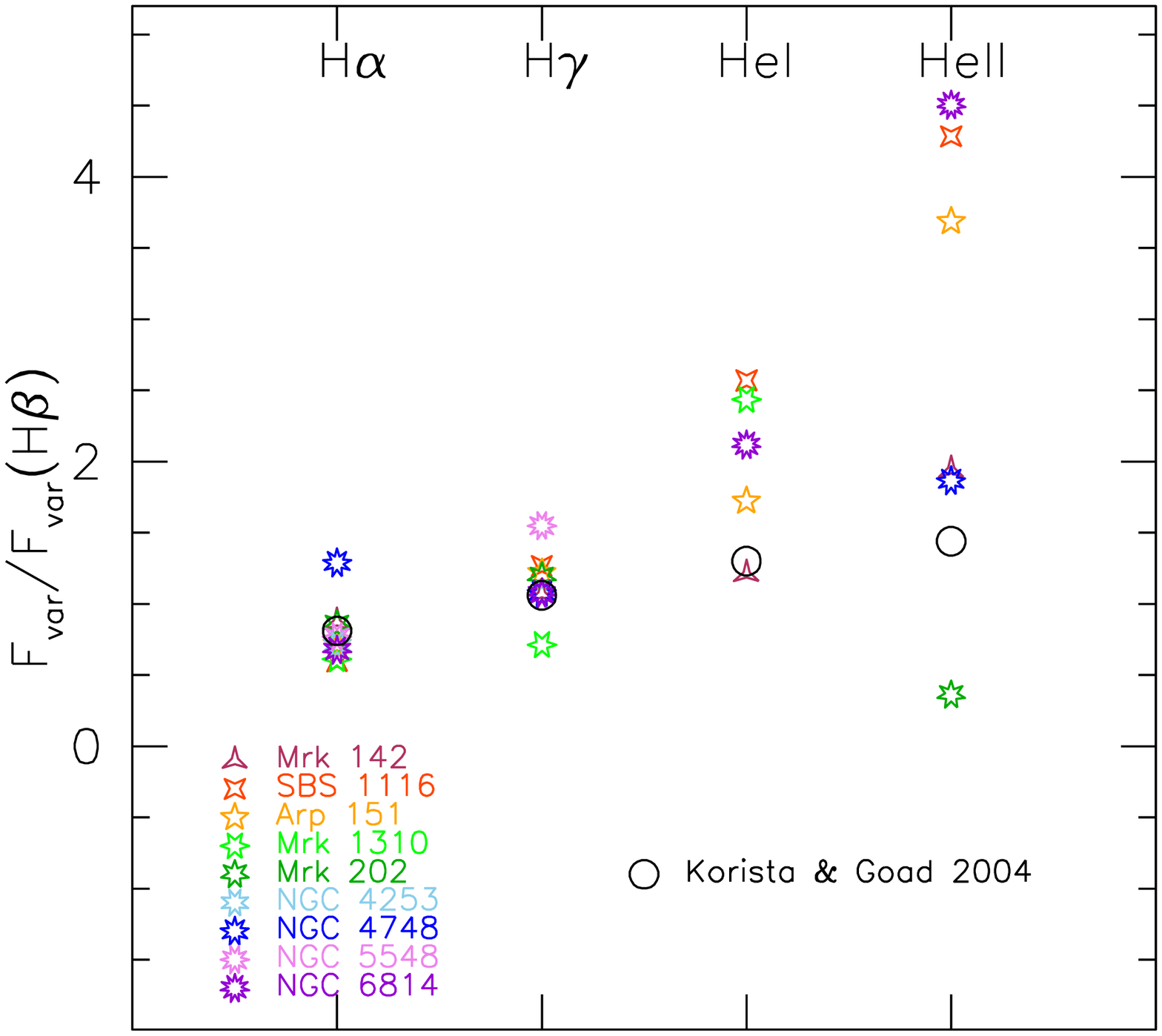}
\caption{{\it Left:} Ratio of the time delay relative to that of
  H$\beta$ for each of the objects and emission lines examined here.
  The weighted mean ratios are found to be $\tau{\rm (H\alpha)} :
  \tau{\rm (H\beta)} : \tau{\rm (H\gamma)} : \tau$(\ion{He}{1}) $ :
  \tau$(\ion{He}{2}) $ = 1.54 : 1.00 : 0.61 : 0.36 : 0.25$.  This is
  in the same direction as the trend predicted by \citet{korista04}
  for the responsivity-weighted radius of emission for the optical
  recombination lines. {\it Right:} Ratio of the excess variance
  relative to that of H$\beta$ for each of the objects and emission
  lines examined here.  The open black circles show the predicted
  emission-line responsivity relative to that of H$\beta$ as tabulated
  by \citet{korista04}.  The trend seen here of $\eta$(\ion{He}{2})$ >
  \eta$(\ion{He}{1})$ > \eta{\rm (H\gamma)} > \eta\rm{ (H\beta)} >
  \eta{\rm (H\alpha)}$ is in the same direction as that predicted by
  \citeauthor{korista04}, although the observed trend is a bit steeper
  than predicted.}
\label{fig:tau.eta}
\end{figure*}

Among the Balmer lines, we find a trend of $\tau{\rm (H\alpha)} >
\tau{\rm (H\beta)} > \tau{\rm (H\gamma)}$.  This trend is seen in
other monitoring studies of multiple AGNs (e.g., \citep{kaspi00}), but
it has not been particularly significant in previous studies due to
the large uncertainties in the time-lag measurements.  Under pure
recombination, the emission from all the Balmer lines would be
expected to originate from the same location in the BLR.  However, it
has long been known that an additional process beyond recombination
must be affecting the observed behavior of the broad Balmer lines in
AGNs, as evidenced by a variable Balmer decrement (e.g.,
\citealt{peterson86,cohen86}).  The modification of pure recombination
effects is theorized to be the result of radial stratification of
optical-depth effects within the BLR (e.g.,
\citealt{netzer75,rees89,korista04}).  In essence, the gas densities
of the line-emitting ``clouds'' are higher at smaller radii (closer to
the black hole), so the relative-flux variations are strongly weighted
by gas at larger radii where the densities and optical depths to line
emission are smaller.  This, together with the fact that at a given
continuum flux, the optical depth of H$\alpha$ is largest, followed by
H$\beta$, and so on through the Balmer series, will serve to make the
emission from each of these lines appear to originate at a different
distance from the source (see Figure 3 of \citealt{korista04}), with
the H$\alpha$ emission appearing to originate at the largest radius
(i.e., largest mean time delay).  The optical depths of \ion{He}{1}
$\lambda 5876$ and \ion{He}{2} $\lambda 4686$ are even smaller than
those of the Balmer lines, with that of \ion{He}{2} being the smallest
of all the lines considered here, causing their responsivity-weighted
radii to be even smaller, as we indeed see.  For all of the
``reliable'' lags measured here, the weighted average time-lag ratios
are $\tau{\rm (H\alpha)} : \tau{\rm (H\beta)} : \tau{\rm (H\gamma)} :
\tau$(\ion{He}{1}) $ : \tau$(\ion{He}{2}) $ = 1.54 : 1.00 : 0.61 :
0.36 : 0.25$ (see Figure~\ref{fig:tau.eta}).

The responsivity of the lines within a specific AGN can also be
compared, where the responsivity ($\eta$) of an emission line is a
measure of the efficiency of the BLR gas in converting a {\it change}
in ionizing flux to line flux.  Examination of the values $F_{\rm
  var}$ and $R_{\rm max}$ in Table~\ref{table:variability} shows that,
in general, $\eta$(\ion{He}{2})$ > \eta$(\ion{He}{1})$ > \eta{\rm
  (H\gamma)} > \eta\rm{ (H\beta)} > \eta{\rm (H\alpha)}$ (see
Figure~\ref{fig:tau.eta}).  Comparison of the light curves in
Figures~\ref{fig:mrk142}--\ref{fig:n6814} also illustrates that
proportionally larger variations are seen in the He lines than the
Balmer lines, in response to changes in the continuum flux.  This
trend is in keeping with the predictions of \citet{korista04} and the
findings of previous monitoring programs (e.g.,
\citealt{peterson86,dietrich93,kollatschny03}).

Finally, we find here that the line width measured in the variable
part of the spectrum is typically narrower than the line width
measured in the mean spectrum.  This trend has been seen in most
previous monitoring programs (see \citealt{peterson04}) and is another
prediction that naturally arises from the photoionization calculations
of \citet{korista04}.  The expectation is that the outer wings of the
lines are generated in the inner BLR, where the gas velocities are
high.  However, the ionization is also higher in the inner BLR, and
the gas responsivity is therefore lower.  Hence, the variability of
the wings of the emission lines will be much lower than that of the
line cores, causing the variable part of the emission line to appear
narrower.

\section{Conclusions}

The LAMP sample of AGNs was originally chosen for spectroscopic
monitoring in order to extend to lower masses the range of black hole
scaling relationships in AGNs.  With the high-quality spectroscopic
dataset obtained at Lick Observatory, we are able to go beyond the
original goals of LAMP and begin to examine the details of the BLR
physics in these AGNs in the following ways:

\begin{itemize}

\item We have presented time-delay measurements and line widths for
  all of the optical H and He recombination lines in the spectra of
  the LAMP sample of AGNs: H$\alpha$, H$\beta$, H$\gamma$, \ion{He}{2}
  $\lambda 4686$, and \ion{He}{1} $\lambda 5876$.

\item Comparisons of the black hole masses determined from multiple
  emission lines are consistent within individual sources, even when
  assuming a single scaling factor $f$.  For at least the optical
  recombination lines, it appears that the scaling factor is not
  heavily dependent on the specific emission line when determining
  black hole masses from reverberation mapping.

\item The time lag versus the line-width measurements for multiple
  emission lines in an individual source are generally consistent with
  a virial relationship ($\tau \propto v^{-2}$).  Virial relationships
  have been seen in other AGNs with high-quality spectroscopic
  monitoring data, upholding the use of reverberation-mapping results
  as a probe of the gravitational influence of the supermassive black
  hole on the BLR gas.

\item For six of the LAMP AGNs, we have examined the velocity-resolved
  time-lag response across the broad H$\alpha$, H$\beta$, and
  H$\gamma$ lines.  In three of the AGNs, we find a significant trend
  in the delay versus the velocity across the line profiles of all
  three Balmer lines.  In the other three AGNs, there is no
  significant trend in delay across the line profile, which may, in
  fact, argue for evidence of circular motions in a Keplerian
  potential.  We are currently investigating whether more detailed
  decompositions of the velocity-resolved time-lag response in these
  objects may be accomplished using the maximum entropy method
  \citep{horne91,horne94}.

\item We are able to confirm several trends in the behavior of the
  broad optical recombination lines that are expected from recent
  photoionization calculations and have also typically been seen in
  other AGN monitoring campaigns.  Specifically, we confirm an
  increase in responsivity and a decrease in the mean time lag as the
  excitation and ionization level for an emission line increases.
  This is manifest as $\tau{\rm (H\alpha)} > \tau{\rm (H\beta)} >
  \tau{\rm (H\gamma)} > \tau$(\ion{He}{1}) $ > \tau$(\ion{He}{2}) and
  $\eta{\rm (H\alpha)} < \eta\rm{ (H\beta)} < \eta{\rm (H\gamma)} <
  \eta$(\ion{He}{1})$ < \eta$(\ion{He}{2}).  Agreement with these
  photoionization calculations argues for optical-depth effects that
  appear to ``fine tune'' the responses of the optical recombination
  lines, as expected under the LOC model for AGN BLRs.

\end{itemize}

Many of the additional predictions of \citet{korista04} for optical
recombination lines in AGN BLRs require multiple monitoring campaigns
of multiple emission lines from a single AGN in different flux states.
The investment of time to examine these predictions is both warranted
and necessary.  The optical recombination line emissivities and
responsivities depend on the local continuum flux (i.e., radius) for a
fixed continuum luminosity, and thus the optical recombination
lines are important to include in quasar tomography for mapping out
the physical parameters of BLR \citep{horne03}.  The recovery of a
velocity-delay map for a single emission line, such as H$\beta$, is a
key goal of reverberation mapping and would allowing insight into the
geometry and kinematics of the BLR.  The simultaneous recovery of
velocity-delay maps for several emission lines could set much stronger
constraints on, and perhaps break degeneracies between, the physical
parameters of the line-emitting gas in the BLR and may usher in yet
another new era of understanding for this spatially unresolved region
in AGNs.

\acknowledgements

We would like to thank the excellent staff and support personnel at
Lick Observatory for their enormous help during our observing run, and
Brad Peterson for helpful conversations and the use of his analysis
software.  We also thank Josh Shiode for his observing help.  This
work was supported by NSF grants AST--0548198 (UC Irvine),
AST--0607485 and AST--0908886 (UC Berkeley), AST--0642621 (UC
Santa Barbara), and AST--0507450 (UC Riverside). The UC Berkeley
researchers also gratefully acknowledge the support of both the Sylvia
\& Jim Katzman Foundation and the TABASGO Foundation for the continued
operation of KAIT.  


\clearpage

\clearpage


\begin{deluxetable}{lccccl}
\tablecolumns{6}
\tablewidth{0pt}
\tablecaption{Object List}
\tablehead{
\colhead{Object} &
\colhead{$\alpha_{2000}$} &
\colhead{$\delta_{2000}$} &
\colhead{$z$} &
\colhead{$A_{B}$\tablenotemark{a}} &
\colhead{Alternate} \\
\colhead{} &
\colhead{(hr min sec)} &
\colhead{($\degr\, \arcmin\, \arcsec$)} &
\colhead{} &
\colhead{(mag)} &
\colhead{Name}}
\startdata

Mrk\,142        & 10 25 31.3 & $+$51 40 35 & 0.04494 & 0.069 &  PG\,1022+519 \\
SBS\,1116+583A  & 11 18 57.7 & $+$58 03 24 & 0.02787 & 0.050 &  \\
Arp\,151        & 11 25 36.2 & $+$54 22 57 & 0.02109 & 0.059 &  Mrk\,40 \\
Mrk\,1310       & 12 01 14.3 & $-$03 40 41 & 0.01941 & 0.133 &  \\
Mrk\,202        & 12 17 55.0 & $+$58 39 35 & 0.02102 & 0.087 &  \\
NGC\,4253       & 12 18 26.5 & $+$29 48 46 & 0.01293 & 0.084 &  Mrk\,766 \\
NGC\,4748       & 12 52 12.4 & $-$13 24 53 & 0.01463 & 0.223 &  \\
NGC\,5548       & 14 17 59.5 & $+$25 08 12 & 0.01718 & 0.088 &  \\
NGC\,6814       & 19 42 40.6 & $-$10 19 25 & 0.00521 & 0.790 &  \\

\enddata
\label{table:objects}
\tablenotetext{a}{The Galactic extinction is based on \citet{schlegel98}.}
\end{deluxetable}


\begin{deluxetable}{lccccc}
\tablecolumns{6}
\tablewidth{0pt}
\tabletypesize{\footnotesize}
\tablecaption{Emission-Line Continuum Windows and Integration Limits}
\tablehead{
\colhead{Object} &
\colhead{Line} &
\multicolumn{2}{c}{Continuum Windows} &
\colhead{Line Limits} &
\colhead{$<f> \pm \sigma_{f}$} \\
\colhead{} &
\colhead{} &
\colhead{(\AA)} &
\colhead{(\AA)} &
\colhead{(\AA)} &
\colhead{($10^{-13}$ erg s$^{-1}$ cm$^{-2}$)}
}
\startdata

Mrk\,142        & H$\alpha$  & $6700-6750$ & $6950-7000$ & $6750-6950$ & $2.62  \pm 0.20$  \\
                & H$\beta$   & $4960-5000$ & $5300-5350$ & $5045-5125$ & $0.928 \pm 0.080$ \\
                & H$\gamma$  & $4400-4450$ & $4580-4630$ & $4480-4580$ & $0.436 \pm 0.043$ \\
& \ion{He}{2} $\lambda 4686$ & $4580-4630$ & $4960-5000$ & $4750-4960$ & $0.70  \pm 0.12$  \\
& \ion{He}{1} $\lambda 5876$ & $5825-5875$ & $6250-6300$ & $6100-6200$ & $0.134 \pm 0.016$ \\
SBS\,1116+583A  & H$\alpha$  & $6600-6650$ & $6950-7000$ & $6650-6850$ & $1.035 \pm 0.071$ \\
                & H$\beta$   & $4875-4925$ & $5200-5250$ & $4925-5055$ & $0.262 \pm 0.028$ \\
                & H$\gamma$  & $4390-4425$ & $4530-4580$ & $4425-4508$ & $0.129 \pm 0.017$ \\
\enddata

\tablecomments{The emission-line fluxes above include the contribution
  from the narrow-line component, which is simply a constant flux
  offset.  In addition, the H$\alpha$ fluxes include the contributions
  from [\ion{N}{2}] $\lambda\lambda$6548,~6583 and the H$\gamma$
  fluxes include the contribution from [\ion{O}{3}] $\lambda$4363.}
\label{table:fluxwind}
\end{deluxetable}

\begin{deluxetable}{ccccc}
\tablecolumns{5}
\tablewidth{0pt}
\tablecaption{Emission-Line Light Curves of Mrk\,142}
\tablehead{
\colhead{HJD} &
\colhead{$f$(H$\alpha$)} &
\colhead{$f$(H$\gamma$)} &
\colhead{$f$(\ion{He}{2})} &
\colhead{$f$(\ion{He}{1})}}

\startdata

 4550.6599	& $2.586  \pm	0.051$  &  $0.436 \pm  0.015$  &  $0.731 \pm  0.030$  &  $0.137 \pm  0.007$ \\
 4551.6560	& $2.448  \pm	0.049$  &  $0.404 \pm  0.014$  &  $0.672 \pm  0.028$  &  $0.116 \pm  0.006$ \\
 4553.6576	& $2.634  \pm	0.052$  &  $0.448 \pm  0.016$  &  $0.680 \pm  0.028$  &  $0.134 \pm  0.007$ \\
 4555.8322	& $3.420  \pm	0.068$  &  $0.566 \pm  0.020$  &  $1.152 \pm  0.048$  &  $0.149 \pm  0.008$ \\
 4556.6591	& $2.581  \pm	0.051$  &  $0.438 \pm  0.015$  &  $0.692 \pm  0.029$  &  $0.126 \pm  0.007$ \\
4557.6574	& $2.575  \pm	0.051$  &  $0.438 \pm  0.015$  &  $0.675 \pm  0.028$  &  $0.141 \pm  0.007$ \\
4558.6464	& $2.679  \pm	0.053$  &  $0.496 \pm  0.017$  &  $0.708 \pm  0.029$  &  $0.145 \pm  0.008$ \\
4559.8879	& $2.524  \pm	0.050$  &  $0.424 \pm  0.015$  &  $0.638 \pm  0.026$  &  $0.118 \pm  0.006$ \\
\enddata

\tablecomments{HJD = Heliocentric Julian Day$-$2,450,000.  Emission-line
  fluxes are in units of $10^{-13}$ erg s$^{-1}$ cm$^{-2}$.  The
  H$\beta$ light curve is presented in Paper~III.}  
\label{table:lc.mrk142}
\end{deluxetable}   


\begin{deluxetable}{ccccc}
\tablecolumns{5}
\tablewidth{0pt}
\tablecaption{Emission-Line Light Curves of SBS\,1116+583A}
\tablehead{
\colhead{HJD} &
\colhead{$f$(H$\alpha$)} &
\colhead{$f$(H$\gamma$)} &
\colhead{$f$(\ion{He}{2})} &
\colhead{$f$(\ion{He}{1})}}

\startdata

 4550.6925   & $11.04 \pm  0.34$   & $1.759  \pm	0.072$   &   $1.215  \pm  0.119$   &   $0.507 \pm  0.099$ \\
 4551.7189   & $11.01 \pm  0.34$   & $1.552  \pm	0.064$   &   $0.834  \pm  0.082$   &   $0.499 \pm  0.097$ \\
 4553.7176   & $10.53 \pm  0.33$   & $1.451  \pm	0.060$   &   $0.661  \pm  0.065$   &   $0.559 \pm  0.109$ \\
 4555.8587   & $11.43 \pm  0.35$   & $1.151  \pm	0.047$   &   $0.830  \pm  0.081$   &   \nodata            \\
 4556.6866   & $10.36 \pm  0.32$   & $1.328  \pm	0.055$   &   $0.431  \pm  0.042$   &   $0.469 \pm  0.091$ \\
4557.6847   & $10.20 \pm  0.32$   & $1.321  \pm	0.054$   &   $0.664  \pm  0.065$   &   $0.365 \pm  0.071$ \\
4558.6633   & $10.41 \pm  0.32$   & $1.393  \pm	0.057$   &   $0.940  \pm  0.092$   &   $0.666 \pm  0.130$ \\
4559.9038   & $11.04 \pm  0.34$   & $1.287  \pm	0.053$   &   $0.388  \pm  0.038$   &   $0.444 \pm  0.087$ \\
\enddata

\tablecomments{HJD = Heliocentric Julian Day$-$2,450,000.  Emission-line
  fluxes are in units of $10^{-14}$ erg s$^{-1}$ cm$^{-2}$.  The
  H$\beta$ light curve is presented in Paper~III.} 

\label{table:lc.sbs}
\end{deluxetable}  


\begin{deluxetable}{ccccc}
\tablecolumns{5}
\tablewidth{0pt}
\tablecaption{Emission-Line Light Curves of Arp\,151}
\tablehead{
\colhead{HJD} &
\colhead{$f$(H$\alpha$)} &
\colhead{$f$(H$\gamma$)} &
\colhead{$f$(\ion{He}{2})} &
\colhead{$f$(\ion{He}{1})}}

\startdata

 4550.7180  &  $3.310  \pm	0.061$  &  $0.318  \pm  0.012$  &  $0.017  \pm  0.004$  &  $0.090  \pm	0.010$ \\
 4551.7478  &  $3.212  \pm	0.059$  &  $0.318  \pm  0.012$  &  $0.023  \pm  0.005$  &  $0.082  \pm	0.009$ \\
 4553.7470  &  $3.031  \pm	0.056$  &  $0.311  \pm  0.012$  &  $0.018  \pm  0.004$  &  $0.085  \pm	0.009$ \\
 4556.7118  &  $3.060  \pm	0.057$  &  $0.247  \pm  0.009$  &  $0.040  \pm  0.009$  &  $0.075  \pm	0.008$ \\
 4557.7555  &  $2.948  \pm	0.055$  &  $0.245  \pm  0.009$  &  $0.020  \pm  0.005$  &  $0.063  \pm	0.007$ \\
4558.6902  &  $2.854  \pm	0.053$  &  $0.248  \pm  0.009$  &  $0.005  \pm  0.001$  &  $0.086  \pm	0.009$ \\
4559.8700  &  $2.813  \pm	0.052$  &  $0.254  \pm  0.010$  &  $0.015  \pm  0.004$  &  $0.066  \pm	0.007$ \\
4560.6570  &  $2.825  \pm	0.052$  &  $0.263  \pm  0.010$  &  $0.030  \pm  0.007$  &  $0.075  \pm	0.008$ \\
\enddata

\tablecomments{HJD = Heliocentric Julian Day$-$2,450,000.  Emission-line
  fluxes are in units of $10^{-13}$ erg s$^{-1}$ cm$^{-2}$.  The
  H$\beta$ light curve is presented in Paper~III.} 

\label{table:lc.arp151}
\end{deluxetable}


\begin{deluxetable}{ccccc}
\tablecolumns{5}
\tablewidth{0pt}
\tablecaption{Emission-Line Light Curves of Mrk\,1310}
\tablehead{
\colhead{HJD} &
\colhead{$f$(H$\alpha$)} &
\colhead{$f$(H$\gamma$)} &
\colhead{$f$(\ion{He}{2})} &
\colhead{$f$(\ion{He}{1})}}

\startdata

 4550.7726	& $2.491  \pm	0.061$  &  $0.288  \pm  0.009$	 & $0.0321  \pm	0.0074$  &  $0.076  \pm	0.009$ \\
 4551.8100	& $2.399  \pm	0.059$  &  $0.321  \pm  0.010$	 & $0.0588  \pm	0.0135$  &  $0.087  \pm	0.010$ \\
 4553.8092	& $2.476  \pm	0.060$  &  $0.339  \pm  0.010$	 & $0.0588  \pm	0.0135$  &  $0.106  \pm	0.012$ \\
 4556.7724	& $2.464  \pm	0.060$  &  $0.354  \pm  0.010$	 & $0.0583  \pm	0.0134$  &  $0.127  \pm	0.014$ \\
 4557.8255	& $2.570  \pm	0.063$  &  $0.347  \pm  0.010$	 & $0.0298  \pm	0.0068$  &  $0.085  \pm	0.010$ \\
4558.7907	& $2.572  \pm	0.063$  &  $0.350  \pm  0.010$	 & $0.0297  \pm	0.0068$  &  $0.106  \pm	0.012$ \\
4559.9485	& $2.643  \pm	0.064$  &  $0.295  \pm  0.009$	 & $0.0002  \pm	0.0005$  &  $0.098  \pm	0.011$ \\
4560.7912	& $2.531  \pm	0.062$  &  $0.332  \pm  0.010$	 & $0.0223  \pm	0.0051$  &  $0.092  \pm	0.010$ \\
\enddata

\tablecomments{HJD = Heliocentric Julian Day$-$2,450,000.  Emission-line
  fluxes are in units of $10^{-13}$ erg s$^{-1}$ cm$^{-2}$.  The
  H$\beta$ light curve is presented in Paper~III.} 

\label{table:lc.mrk1310}
\end{deluxetable}


\begin{deluxetable}{ccccc}
\tablecolumns{5}
\tablewidth{0pt}
\tablecaption{Emission-Line Light Curves of Mrk\,202}
\tablehead{
\colhead{HJD} &
\colhead{$f$(H$\alpha$)} &
\colhead{$f$(H$\gamma$)} &
\colhead{$f$(\ion{He}{2})} &
\colhead{$f$(\ion{He}{1})}}

\startdata

 4550.7434  & $1.292  \pm  0.023$  &  $0.1489  \pm  0.0048$  &  $0.1605  \pm  0.0088$  &  $0.0489  \pm  0.0102$ \\
 4551.7693  & $1.247  \pm  0.022$  &  $0.1574  \pm  0.0051$  &  $0.1420  \pm  0.0078$  &  $0.0421  \pm  0.0088$ \\
 4553.7721  & $1.339  \pm  0.024$  &  $0.1629  \pm  0.0052$  &  $0.1707  \pm  0.0093$  &  $0.0518  \pm  0.0108$ \\
 4556.7381  & $1.242  \pm  0.022$  &  $0.1439  \pm  0.0046$  &  $0.1056  \pm  0.0058$  &  $0.0476  \pm  0.0099$ \\
 4557.7884  & $1.259  \pm  0.022$  &  $0.1460  \pm  0.0047$  &  $0.1066  \pm  0.0058$  &  $0.0200  \pm  0.0042$ \\
4558.7121  & $1.262  \pm  0.022$  &  $0.1286  \pm  0.0041$  &  $0.1032  \pm  0.0056$  &  $0.0423  \pm  0.0088$ \\
4559.9658  & $1.421  \pm  0.025$  &  $0.1224  \pm  0.0039$  &  $0.1028  \pm  0.0056$  &  $0.0677  \pm  0.0141$ \\
4560.7389  & $1.241  \pm  0.022$  &  $0.1432  \pm  0.0046$  &  $0.1212  \pm  0.0066$  &  $0.0429  \pm  0.0089$ \\
\enddata

\tablecomments{HJD = Heliocentric Julian Day$-$2,450,000.  Emission-line
  fluxes are in units of $10^{-13}$ erg s$^{-1}$ cm$^{-2}$. The
  H$\beta$ light curve is presented in Paper~III.} 

\label{table:lc.mrk202}
\end{deluxetable}


\begin{deluxetable}{ccccc}
\tablecolumns{5}
\tablewidth{0pt}
\tablecaption{Emission-Line Light Curves of NGC\,4253}
\tablehead{
\colhead{HJD} &
\colhead{$f$(H$\alpha$)} &
\colhead{$f$(H$\gamma$)} &
\colhead{$f$(\ion{He}{2})} &
\colhead{$f$(\ion{He}{1})}}

\startdata

 4550.8014  &  $10.57  \pm  0.25$  &  $0.623  \pm  0.017$  &  $0.628  \pm  0.028$  &  $0.282  \pm  0.011$ \\
 4551.8338  &  $10.63  \pm  0.25$  &  $0.653  \pm  0.018$  &  $0.642  \pm  0.029$  &  $0.299  \pm  0.012$ \\
 4553.8298  &  $10.61  \pm  0.25$  &  $0.697  \pm  0.019$  &  $0.729  \pm  0.033$  &  $0.295  \pm  0.011$ \\
 4555.9872  &  $10.73  \pm  0.25$  &  $0.448  \pm  0.074$  &  $0.590  \pm  0.027$  &  $0.298  \pm  0.052$ \\
 4556.7926  &  $10.60  \pm  0.25$  &  $0.690  \pm  0.019$  &  $0.694  \pm  0.031$  &  $0.304  \pm  0.012$ \\
4557.7073  &  $11.59  \pm  0.27$  &  $0.688  \pm  0.019$  &  $0.683  \pm  0.031$  &  $0.328  \pm  0.013$ \\
4558.6802  &  $11.24  \pm  0.26$  &  $0.685  \pm  0.019$  &  $0.674  \pm  0.030$  &  $0.345  \pm  0.013$ \\
4559.9827  &  $11.74  \pm  0.28$  &  $0.719  \pm  0.020$  &  $0.644  \pm  0.029$  &  $0.329  \pm  0.013$ \\
\enddata

\tablecomments{HJD = Heliocentric Julian Day$-$2,450,000.  Emission-line
  fluxes are in units of $10^{-13}$ erg s$^{-1}$ cm$^{-2}$.  The
  H$\beta$ light curve is presented in Paper~III.} 

\label{table:lc.n4253}
\end{deluxetable}


\begin{deluxetable}{ccccc}
\tablecolumns{5}
\tablewidth{0pt}
\tablecaption{Emission-Line Light Curves of NGC\,4748}
\tablehead{
\colhead{HJD} &
\colhead{$f$(H$\alpha$)} &
\colhead{$f$(H$\gamma$)} &
\colhead{$f$(\ion{He}{2})} &
\colhead{$f$(\ion{He}{1})}}

\startdata

 4550.8173  & $8.65  \pm  0.19$  &  $0.965  \pm  0.023$  &  $1.029  \pm  0.036$  &  $0.329  \pm  0.031$ \\
 4551.8530  & $8.58  \pm  0.19$  &  $1.037  \pm  0.025$  &  $1.004  \pm  0.035$  &  $0.293  \pm  0.027$ \\
 4553.8454  & $8.59  \pm  0.19$  &  $1.040  \pm  0.025$  &  $1.006  \pm  0.035$  &  $0.294  \pm  0.028$ \\
 4556.8087  & $8.31  \pm  0.18$  &  $0.979  \pm  0.023$  &  $1.007  \pm  0.035$  &  $0.342  \pm  0.032$ \\
 4557.8618  & $8.50  \pm  0.19$  &  $0.977  \pm  0.023$  &  $0.888  \pm  0.031$  &  $0.295  \pm  0.028$ \\
4558.8076  & $8.60  \pm  0.19$  &  $1.019  \pm  0.024$  &  $0.947  \pm  0.033$  &  $0.324  \pm  0.030$ \\
4560.8578  & $8.52  \pm  0.19$  &  $1.029  \pm  0.025$  &  $1.112  \pm  0.039$  &  $0.341  \pm  0.032$ \\
4561.8522  & $8.81  \pm  0.19$  &  $1.016  \pm  0.024$  &  $1.131  \pm  0.039$  &  $0.362  \pm  0.034$ \\
\enddata

\tablecomments{HJD = Heliocentric Julian Day$-$2,450,000.  Emission-line
  fluxes are in units of $10^{-13}$ erg s$^{-1}$ cm$^{-2}$.  The
  H$\beta$ light curve is presented in Paper~III.} 

\label{table:lc.n4748}
\end{deluxetable}


\begin{deluxetable}{ccccc}
\tablecolumns{5}
\tablewidth{0pt}
\tablecaption{Emission-Line Light Curves of NGC\,5548}
\tablehead{
\colhead{HJD} &
\colhead{$f$(H$\alpha$)} &
\colhead{$f$(H$\gamma$)} &
\colhead{$f$(\ion{He}{2})} &
\colhead{$f$(\ion{He}{1})}}

\startdata

 4550.8678  &  $17.04  \pm  0.24$  &  $0.802  \pm  0.034$  &  $0.326  \pm  0.054$  &  $1.56  \pm  0.11$ \\
 4551.8664  &  $16.97  \pm  0.24$  &  $0.777  \pm  0.033$  &  $0.277  \pm  0.046$  &  $1.73  \pm  0.12$ \\
 4553.8626  &  $16.81  \pm  0.23$  &  $0.796  \pm  0.034$  &  $0.283  \pm  0.047$  &  $1.61  \pm  0.11$ \\
 4556.8589  &  $16.98  \pm  0.24$  &  $0.753  \pm  0.032$  &  $0.352  \pm  0.058$  &  $1.56  \pm  0.10$ \\
 4557.8755  &  $16.80  \pm  0.23$  &  $0.795  \pm  0.034$  &  $0.270  \pm  0.045$  &  $1.56  \pm  0.11$ \\
4558.8595  &  $16.92  \pm  0.24$  &  $0.758  \pm  0.032$  &  $0.264  \pm  0.044$  &  $1.49  \pm  0.10$ \\
4559.9903  &  $18.25  \pm  0.25$  &  $0.738  \pm  0.032$  &  $0.205  \pm  0.034$  &  $1.51  \pm  0.10$ \\
4560.8967  &  $16.29  \pm  0.23$  &  $0.794  \pm  0.034$  &  $0.292  \pm  0.048$  &  $1.50  \pm  0.10$ \\
\enddata

\tablecomments{HJD = Heliocentric Julian Day$-$2,450,000.  Emission-line
  fluxes are in units of $10^{-13}$ erg s$^{-1}$ cm$^{-2}$.  The
  H$\beta$ light curve is presented in Paper~III.} 

\label{table:lc.n5548}
\end{deluxetable}


\begin{deluxetable}{ccccc}
\tablecolumns{5}
\tablewidth{0pt}
\tablecaption{Emission-Line Light Curves of NGC\,6814}
\tablehead{
\colhead{HJD} &
\colhead{$f$(H$\alpha$)} &
\colhead{$f$(H$\gamma$)} &
\colhead{$f$(\ion{He}{2})} &
\colhead{$f$(\ion{He}{1})}}

\startdata

 4551.0180  & $11.79  \pm  0.15$  &  $0.937  \pm  0.026$  &  $0.593  \pm  0.126$  &  $0.381  \pm  0.027$ \\
 4552.0264  & $11.13  \pm  0.14$  &  $0.900  \pm  0.025$  &  $0.163  \pm  0.035$  &  $0.527  \pm  0.038$ \\
 4556.0339  & $11.60  \pm  0.15$  &  $0.934  \pm  0.033$  &  $0.457  \pm  0.097$  &  $0.501  \pm  0.036$ \\
 4557.0200  & $12.15  \pm  0.15$  &  $1.042  \pm  0.029$  &  $0.640  \pm  0.136$  &  $0.566  \pm  0.041$ \\
 4560.0402  & $12.66  \pm  0.16$  &  $1.094  \pm  0.038$  &  $0.902  \pm  0.192$  &  $0.892  \pm  0.064$ \\
4561.0266  & $11.93  \pm  0.15$  &  $1.137  \pm  0.032$  &  $0.825  \pm  0.176$  &  $0.634  \pm  0.046$ \\
4564.0146  & $12.31  \pm  0.16$  &  $1.172  \pm  0.033$  &  $0.897  \pm  0.191$  &  $0.650  \pm  0.047$ \\
4567.0021  & $13.63  \pm  0.17$  &  $1.132  \pm  0.032$  &  $0.576  \pm  0.123$  &  $0.640  \pm  0.046$ \\
\enddata

\tablecomments{HJD = Heliocentric Julian Day$-$2,450,000.  Emission-line
  fluxes are in units of $10^{-13}$ erg s$^{-1}$ cm$^{-2}$.  The
  H$\beta$ light curve is presented in Paper~III.} 

\label{table:lc.n6814}
\end{deluxetable}

\begin{deluxetable}{lccccccc}
\tablecolumns{8}
\tablewidth{0pt}
\tablecaption{Light-Curve Statistics}
\tablehead{
\colhead{Object} &
\colhead{Time Series} &
\colhead{$N$} &
\colhead{$\langle T \rangle$} &
\colhead{$T_{\rm median}$} &
\colhead{$\langle \sigma_{\rm f}/f \rangle$} &
\colhead{$F_{\rm var}$} &
\colhead{$R_{\rm max}$}\\
\colhead{(1)} &
\colhead{(2)} &
\colhead{(3)} &
\colhead{(4)} &
\colhead{(5)} &
\colhead{(6)} &
\colhead{(7)} &
\colhead{(8)} }
\startdata

Mrk\,142       & $B$         & 64 & $1.8 \pm 2.3$ & 1.02 & 0.0166 & 0.025 & $1.15 \pm 0.03$ \vspace{0.00in}\\
	       & $V$         & 62 & $1.7 \pm 2.0$ & 1.02 & 0.0119 & 0.024 & $1.12 \pm 0.02$ \vspace{0.00in}\\
               & 5100\,\AA   & 51 & $1.4 \pm 1.0$ & 1.00 & 0.0115 & 0.090 & $1.86 \pm 0.03$ \vspace{0.00in}\\
	       & H$\alpha$   & 50 & $1.4 \pm 1.0$ & 1.00 & 0.0199 & 0.075 & $1.86 \pm 0.05$ \vspace{0.00in}\\
	       & H$\beta$    & 51 & $1.4 \pm 1.0$ & 1.00 & 0.0113 & 0.086 & $1.97 \pm 0.03$ \vspace{0.00in}\\
	       & H$\gamma$   & 51 & $1.4 \pm 1.0$ & 1.00 & 0.0350 & 0.092 & $2.04 \pm 0.10$ \vspace{0.00in}\\
& \ion{He}{2} $\lambda 4686$ & 51 & $1.4 \pm 1.0$ & 1.00 & 0.0414 & 0.167 & $2.64 \pm 0.16$ \vspace{0.00in}\\
& \ion{He}{1} $\lambda 5876$ & 50 & $1.4 \pm 1.0$ & 1.00 & 0.0530 & 0.104 & $1.93 \pm 0.15$ \vspace{0.10in}\\
\enddata 

\tablecomments{Columns are presented as follows: (1) object; (2)
               feature; (3) number of observations; (4) average
               interval (days) between observations; (5) median
               sampling rate (days); (6) mean fractional error; (7)
               excess variance as described in the text; and (8) the
               ratio of the maximum to the minimum flux.}
\label{table:variability}
\end{deluxetable}


\clearpage
\begin{landscape}
\LongTables
\begin{deluxetable}{lcccccccccccc}
\tablecolumns{13}
\tablewidth{0pt}
\tabletypesize{\scriptsize}
\setlength{\tabcolsep}{0.02in} 
\tablecaption{Time-Lag Measurements}
\tablehead{
\colhead{} &
\colhead{} &
\multicolumn{5}{c}{vs. $B$ band} &
\colhead{} &
\multicolumn{5}{c}{vs. $V$ band} \\
\cline{3-7}
\cline{9-13}
\colhead{} &
\colhead{} &
\multicolumn{2}{c}{Observed} &
\multicolumn{2}{c}{Rest frame} &
\colhead{} &
\colhead{} &
\multicolumn{2}{c}{Observed} &
\multicolumn{2}{c}{Rest frame} &
\colhead{} \\ 
\colhead{Object} &
\colhead{Line} &
\colhead{$\tau_{\rm cent}$} &
\colhead{$\tau_{\rm peak}$} &
\colhead{$\tau_{\rm cent}$} &
\colhead{$\tau_{\rm peak}$} &
\colhead{$r_{\rm max}$} &
\colhead{} &
\colhead{$\tau_{\rm cent}$} &
\colhead{$\tau_{\rm peak}$} &
\colhead{$\tau_{\rm cent}$} &
\colhead{$\tau_{\rm peak}$} &
\colhead{$r_{\rm max}$} \\
\colhead{} &
\colhead{} &
\colhead{(days)} &
\colhead{(days)} &
\colhead{(days)} &
\colhead{(days)} &
\colhead{} &
\colhead{} &
\colhead{(days)} &
\colhead{(days)} &
\colhead{(days)} &
\colhead{(days)} &
\colhead{}} 
\startdata

Mrk\,142       	& H$\alpha$    & $2.90^{+1.22}_{-0.92}$ & $2.58 ^{+1.75}_{-0.50}$ & $2.78 ^{+1.17}_{-0.88}$ & $2.47 ^{+1.67}_{-0.48}$ & $0.53 \pm 0.08$ && $2.63 ^{+1.24 }_{-1.02	}$ & $2.50 ^{+1.50 }_{-1.00 }$ & $2.52 ^{+1.18 }_{-0.98}$ & $2.39 ^{+1.44 }_{-0.96 }$ & $0.56 \pm 0.08$ \vspace{0.05in} \\             
                & H$\beta$     & $2.87^{+0.76}_{-0.87}$ & $2.75 ^{+1.00}_{-0.75}$ & $2.74 ^{+0.73}_{-0.83}$ & $2.63 ^{+0.96}_{-0.72}$ & $0.52 \pm 0.07$ && $2.88 ^{+1.00 }_{-1.01	}$ & $3.25 ^{+0.75 }_{-1.75 }$ & $2.76 ^{+0.96 }_{-0.96}$ & $3.11 ^{+0.72 }_{-1.67 }$ & $0.55 \pm 0.08$ \vspace{0.05in} \\ 
	        & H$\gamma$    & $2.99^{+1.27}_{-1.10}$ & $2.75 ^{+1.75}_{-1.00}$ & $2.86 ^{+1.22}_{-1.05}$ & $2.63 ^{+1.67}_{-0.96}$ & $0.54 \pm 0.08$ && $3.08 ^{+1.55 }_{-1.20	}$ & $3.25 ^{+1.50 }_{-1.75 }$ & $2.95 ^{+1.48 }_{-1.15}$ & $3.11 ^{+1.44 }_{-1.67 }$ & $0.58 \pm 0.07$ \vspace{0.05in} \\ 
  & \ion{He}{2} $\lambda 4686$ & $1.25^{+1.60}_{-2.08}$ & $1.25 ^{+1.50}_{-2.00}$ & $1.20 ^{+1.53}_{-1.99}$ & $1.20 ^{+1.44}_{-1.91}$ & $0.50 \pm 0.07$ && $1.61 ^{+1.11 }_{-1.48	}$ & $1.50 ^{+1.25 }_{-1.25 }$ & $1.54 ^{+1.06 }_{-1.42}$ & $1.44 ^{+1.20 }_{-1.20 }$ & $0.56 \pm 0.08$ \vspace{0.05in} \\ 
  & \ion{He}{1}	$\lambda 5876$ & $1.89^{+2.08}_{-1.26}$ & $2.00 ^{+2.25}_{-1.50}$ & $1.81 ^{+1.99}_{-1.20}$ & $1.91 ^{+2.15}_{-1.44}$ & $0.50 \pm 0.06$ && $1.74 ^{+3.77 }_{-1.61	}$ & $1.75 ^{+3.75 }_{-1.50 }$ & $1.66 ^{+3.61 }_{-1.54}$ & $1.67 ^{+3.59 }_{-1.44 }$ & $0.52 \pm 0.07$ \vspace{0.15in} \\	      

SBS\,1116+583A	& H$\alpha$    & $4.12^{+1.41}_{-0.98}$ & $3.75 ^{+1.25}_{-0.75}$ & $4.01 ^{+1.37}_{-0.95}$ & $3.65 ^{+1.22}_{-0.73}$ & $0.63 \pm 0.07$ && $3.34 ^{+0.67 }_{-0.85}$ & $3.25 ^{+0.75 }_{-1.00 }$ & $3.25 ^{+0.65 }_{-0.83}$ & $3.16 ^{+0.73 }_{-0.97 }$ & $0.56 \pm 0.08$ \vspace{0.05in} \\
\enddata
\label{table:tau}
\tablecomments{H$\beta$ lags were first presented in Paper III and are
  included here for reference.  Lag values preceded by a colon are
  considered unreliable (see the text for details).}
\end{deluxetable}
\clearpage
\end{landscape}


\begin{deluxetable}{lccccc}
\tablecolumns{6}
\tablewidth{0pt}
\tablecaption{Rest-Frame Broad Emission Line-Width Measurements}
\tablehead{
\colhead{} &
\colhead{} &
\multicolumn{2}{c}{Mean Spectrum} &
\multicolumn{2}{c}{RMS Spectrum} \\
\colhead{Object} &
\colhead{Line} &
\colhead{$\sigma_{\rm line}$} &
\colhead{FWHM} &
\colhead{$\sigma_{\rm line}$} &
\colhead{FWHM}\\
\colhead{} &
\colhead{} &
\colhead{(km s$^{-1}$)} &
\colhead{(km s$^{-1}$)} &
\colhead{(km s$^{-1}$)} &
\colhead{(km s$^{-1}$)} }
\startdata

Mrk\,142        & H$\alpha$    & $925  \pm 38$ & $1350 \pm 39$    & $934 \pm 61$   & $1262 \pm 166$   \\
                & H$\beta$     & $1116 \pm 22$ & $1462 \pm 2$     & $859 \pm 102$  & $1368 \pm 379$   \\
                & H$\gamma$    & $1114 \pm 67$ & $1889 \pm 69$    & $1057 \pm 82$  & $2547 \pm 533$   \\
  & \ion{He}{2} $\lambda 4686$ & \nodata       & \nodata          & $3079 \pm 166$ & $8061 \pm 1325$  \\ 
  & \ion{He}{1}	$\lambda 5876$ & \nodata       & \nodata          & $1128 \pm 241$ & $5703 \pm 3440$ \vspace{0.15in} \\

SBS\,1116+583A  & H$\alpha$    & $1250^{+50}_{-20}$ & $2059^{+508}_{-16}$ & $1218^{+147}_{-99}$ & $1965^{+1408}_{-374}$ \\
                & H$\beta$     & $1552 \pm 36$ & $3668 \pm 186$   & $1528 \pm 184$ & $3604 \pm  1123$ \\
                & H$\gamma$    & $1084 \pm 33$ & $3877 \pm 98$    & $1402 \pm 95$  & $4241 \pm  1465$ \\
\enddata \tablecomments{H$\beta$ line widths were first presented in
  Paper III and are included here for reference.  Line widths preceded
  by a colon are considered unreliable (see the text for details).}
\label{table:width}
\end{deluxetable}


\begin{deluxetable}{lccc}
\tablecolumns{4}
\tablewidth{0pt}
\tablecaption{Virial Products and Derived Black Hole Masses}
\tablehead{
\colhead{Object} &
\colhead{Line} &
\colhead{$c \tau_{\rm cent} \sigma^2_{\rm line} / G$} &
\colhead{\mbh\tablenotemark{a}}\\
\colhead{} &
\colhead{} &
\colhead{($10^6$ M$_{\odot}$)} &
\colhead{($10^6$ M$_{\odot}$)} }
\startdata

Mrk\,142        & H$\alpha$    & $0.47^{+0.21}_{-0.16}$ & $2.48^{+1.09}_{-0.85}$ \vspace{0.05in}\\
                & H$\beta$     & $0.40^{+0.14}_{-0.15}$ & $2.07^{+0.74}_{-0.80}$ \vspace{0.05in}\\ 
                & H$\gamma$    & $0.62^{+0.28}_{-0.25}$ & $3.3 ^{+1.5 }_{-1.3 }$ \vspace{0.05in}\\
  & \ion{He}{2} $\lambda 4686$ & $2.2 ^{+2.8 }_{-3.7 }$ & $12  ^{+15  }_{-19  }$ \vspace{0.05in}\\
  & \ion{He}{1}	$\lambda 5876$ & $0.45^{+0.53}_{-0.36}$ & $2.4 ^{+2.8 }_{-1.9 }$ \vspace{0.15in}\\
                                
SBS\,1116+583A  & H$\alpha$    & $1.16^{+0.49}_{-0.33}$ & $6.1 ^{+2.5}_{-1.8}$ \vspace{0.05in}\\
                & H$\beta$     & $1.05^{+0.38}_{-0.34}$ & $5.5 ^{+2.0}_{-1.8}$ \vspace{0.05in}\\ 
                & H$\gamma$    & $0.71^{+0.25}_{-0.22}$ & $3.7 ^{+1.3}_{-1.1}$ \vspace{0.05in}\\
\enddata

\tablenotetext{a}{Assuming $f=5.2$, from \citet{woo10}.  Note that the
  derived masses would increase by $\sim 5\%$ using $f=5.5$ from
  \citet{onken04}.}  
\tablecomments{H$\beta$-based masses were first presented in Paper III
  and are included here for reference.  Values preceded by a colon
  are considered unreliable (see the text for details).}
\label{table:mbh}
\end{deluxetable}

\end{document}